\documentclass[final,5p,times,twocolumn,authoryear]{elsarticle-modif}

\usepackage{graphicx}
\usepackage{bm}
\usepackage{amssymb}
\usepackage{nicefrac}
\usepackage{amsmath}
\usepackage{hyperref}
\usepackage{hypernat}
\usepackage{color}
\newcommand{\St}{\mbox{\textit{St}}}
\newcommand{\Sone}{S_{\!1}}
\newcommand{\Ku}{\mbox{\textit{Ku}}}

\newcommand{\taup}{\tau_\mathrm{p}}

\newcommand{\Ld}{\mathcal{L}_2}
\newcommand{\dv}{\Delta v}
\newcommand{\dx}{\Delta x}
\newcommand{\dt}{\Delta t}
\newcommand{\D}{\mathcal{D}}

\newcommand{\avg}[1]{\langle #1 \rangle}

\begin{document}

%%%

\begin{frontmatter}
  \title{A lattice model for the Eulerian description of heavy
    particle suspensions} \author{Fran\c{c}ois
    Laenen}\ead{francois.laenen@oca.eu} \author{Giorgio
    Krstulovic}\ead{giorgio.krstulovic@oca.eu} \author{J\'{e}r\'{e}mie
    Bec}\ead{jeremie.bec@oca.eu} \cortext[corres]{Corresponding
    author} \address{Laboratoire J.L.\ Lagrange, Universit\'{e} C\^ote
    d'Azur, Observatoire de la C\^{o}te d'Azur, CNRS, Bd.\ de
    l'Observatoire, 06300, Nice, France.}

								%% ABSTRACT %%
																
\begin{abstract}
  Modeling dispersed solid phases in fluids still represents a
  computational challenge when considering a small-scale coupling in
  wide systems, such as the atmosphere or industrial
  processes at high Reynolds numbers. A numerical method is here
  introduced for simulating the dynamics of diffusive heavy inertial
  particles in turbulent flows. The approach is based on the
  position/velocity phase-space particle distribution. The
  discretization of velocities is inspired from lattice Boltzmann
  methods
  and is chosen to match discrete displacements between two time
  steps. For each spatial position, the time evolution of particles
  momentum is approximated by a finite-volume approach. The proposed
  method is tested for particles experiencing a Stokes viscous drag
  with a prescribed fluid velocity field in one dimension using a
  random flow, and in two dimensions with the solution to the
  forced incompressible Navier--Stokes equations. Results show good
  agreements between Lagrangian and Eulerian dynamics for
  both spatial clustering and the dispersion in particle velocities.
  This demonstrates the suitability of the proposed approach at large
  Stokes numbers or for situations where details of collision
  processes are important.
\end{abstract}


\end{frontmatter}

								%% INTRODUCTION %%

\section{Introduction}

Particle-laden turbulent flows are found in numerous natural and
industrial situations, ranging from droplet growth in clouds and dust
accretion in early stellar systems, to turbulent mixing in engines and
industrial sprays. In such situations, the processes that need being
modeled and quantified involve the fine-scale dynamical properties of
the particles, like preferential concentration, collisions and
coalescences, chemical reactions, and modulation of the fluid flow by
the particles. To address specific microphysical issues, one usually
study simultaneously the turbulent flow and the dispersed phase using
Eulerian-Lagrangian direct numerical simulation \citep[see, for
instance,][for recent
reviews]{toschi2009lagrangian,balachandar2010turbulent}. This approach
is particularly suited for monitoring the fluctuations occurring at
dissipative scales. However, direct numerical simulations are quickly
too computationally expensive for studying particle suspensions in
realistic settings. On the one hand, a large-scale system, such as a
chemical reactor, an atmospheric cloud or a protoplanetary disk,
contains a prohibitively large number of particles. On the other hand,
the substantial Reynolds numbers of natural and industrial flows
require the use of large-scale models, such as large-eddy simulations.
Eulerian-Lagrangian methods, where the dispersed phase is modeled by
point particles, show some advantages: they allow for an easy
implementation of polydispersity and are rather insensitive to
subgrid-scale fluctuations, at least for particles with a large-enough
response time \citep{wang1996large}. For particles with smaller
inertia, one relies on the use of stochastic Langevin models
\citep{shotorban2006stochastic,pozorski2009filtered}. In addition the
constraints on the number of particles can be relaxed using
super-particles approaches, which then necessitates some modeling for
collisions \citep{shima2009super,rein2010validity}. However, as
stressed for instance by \cite{portela2006possibilities}, Lagrangian
methods prove some difficulties in correctly predicting modifications
of the carrier flow by the dispersed phase, particle-to-particle
interactions and near-wall effects.

Some of these shortcomings can be circumvent using Eulerian-Eulerian
methods \citep[see][for a review]{fox2012large}. The main difficulty
then relies in finding a fluid description of the dispersed
particulate phase. In principle, this is achieved by prescribing a
closure for the kinetic hierarchy of moment equations. When
considering an ensemble of particle trajectories
$(\bm x_\mathrm{p},\bm v_\mathrm{p})$, one naturally introduces the
phase-space density \begin{equation}
  f(\bm x,\bm v, t) = \left\langle{\delta(\bm
      x_\mathrm{p}(t) - \bm x)\,\delta(\bm v_\mathrm{p}(t)-\bm
      v)}\right\rangle,
  \label{eq:singularDistr}
\end{equation}
where the fluid velocity realization is fixed and the average
$\langle\cdot\rangle$ is both over the particle ensemble (different
realizations of the initial conditions and/or average over a large
number of particles) and over the molecular diffusion of the particles
(with diffusion constant $\kappa$). The phase-space density then
solves the diffusion-transport equation
\begin{equation}
  \partial_t f + \bm v\cdot \nabla_{\bm x}f - \nabla_{\bm v} \cdot
  \left[f\,\mathcal{F}(\bm x, \bm v, t)/m_\mathrm{p}\right] -
  \nabla_{\bm v}\cdot\left(\kappa \nabla_{\bm v} f\right) = 0,
  \label{eq:conslaw}
\end{equation}
where $\mathcal{F}$ is the force exerted by the fluid on a particle
located at $\bm x$ with a velocity $\bm v$ and $m_\mathrm{p}$ is the
particle mass. This Fokker--Planck equation is exact and fully
describes the dynamics of small particles in the phase space. The
drift terms are completely prescribed by a given realization of the
fluid flow. To obtain an Eulerian description of particles dynamics
that depends on the spatial variable $\bm x$ only, the usual approach
consists in deriving the equations for the various moments of the
velocity $\bm v$. To close the resulting hierarchy, additional
assumptions are needed. They naturally arise when focusing on given
asymptotics \citep[see, e.g.,][]{carrillo2008simulation}. For
instance, when the particles experience a very strong viscous drag
with the flow (small Stokes numbers), an effective particle velocity
can be written \citep{maxey1987gravitational} leading to close this
hierarchy at the zeroth order and to write an explicit equation for
the transport of particle density. This then serves as a basis for
deriving subgrid-scale models for large-eddy simulations \citep[see,
e.g., ][]{Shotorban:2007jo}. First-order closures lead to writing an
evolution equation for a particle velocity field that is coupled to
the fluid flow. Again, such methods are limited to asymptotically
small values of the Stokes number, as they are inadequate to deal with
multi-streamed particle distributions. It is indeed well known that
the trajectories of finite-Stokes-number particles can cross, leading
to the formation of regions where the particle velocities are
multivalued and cannot be described in terms of a spatial field. This
phenomenon is usually referred to as \emph{particle-trajectory
  crossing} \citep{chen2000heavy}, \emph{sling
  effect} \citep{falkovich2002acceleration} or \emph{caustic
  formation} \citep{wilkinson2005caustics} and has important impacts
in estimating collision rates \citep[see, e.g.,][]{bec2005clustering}.
Higher-order closures, such as ten-moment equations, can also be
proposed depending on the specific forces applied on the particles.
They account for the dispersion in particle velocities and can thus
catch some aspects of multi-streaming \citep[using either algebraic or
quadrature closures][]{fevrier2005partitioning,desjardins2008quadrature}.

In principle, accessing the full multi-streaming dynamics of particles
requires solving the kinetic equation (\ref{eq:conslaw}) in the entire
$(2\times d)$-dimensional position-velocity phase space. ($d$ denotes
here the dimension of the physical space.) A clear difficulty which is
then faced is the prohibitive computational cost of integrating a
partial differential equation in a space with such a large
dimensionality. Attempts have nevertheless been made by decreasing the
number of relevant degrees of freedom. This can be easily done, for
instance, by considering one-dimensional flows \citep[see, e.g.,][and
references therein]{carrillo2008simulation}. Other approaches are
based on the physical observation that the distribution of particle
velocities is usually rather concentrated along a given number of
branches in phase space. This led for instance \cite{liu2011level} to
capture implicitly the velocity dispersion by applying a level-set
method in phase space. The efficient implementation of this procedure
to high-dimensional turbulent situations still represents a real
challenge.

Here, we propose an alternative approach that consists in degrading
the resolution in velocities and to apply computationally efficient
ideas inspired from Lattice-Boltzmann methods \citep{chen1998lattice}.
The discrete values of velocities are chosen to correspond exactly to
discrete displacements between two time steps on a fixed spatial
lattice. The time evolution of $f({\bm x},{\bm v},t)$ is then
approximated by splitting the spatial advection on the lattice and the
acceleration of particles, which is integrated using a finite-volume
scheme. This gives access to a full phase-space particle distribution
that naturally catch multi-streaming. This method is relevant to cases
where diffusion is responsible for a broadening of the particle
velocity dispersion and it applies to any kind of force $\mathcal{F}$
acting on the particles. After describing the algorithm in
Sec.~\ref{sec:AlgDescr}, we present some qualitative and quantitative
tests for very small heavy particles whose dynamics is dominated by
diffusion and a Stokes drag with the fluid velocity. The proposed
lattice-particle method is directly compared to the results of direct
Lagrangian simulations. Section~\ref{sec:1D} is devoted to the
one-dimensional case with a random Gaussian flow. Section~\ref{sec:2D}
shows the results of coupled Navier--Stokes and lattice particles
simulations in two dimensions for turbulent flows, either in the
direct cascade of enstrophy or in the inverse cascade of kinetic
energy.

%% SECTION ALG Desc%%

\section{Description of the method}
\label{sec:AlgDescr}

\begin{figure}[ht]
  \centerline{\includegraphics[width=0.7\columnwidth]{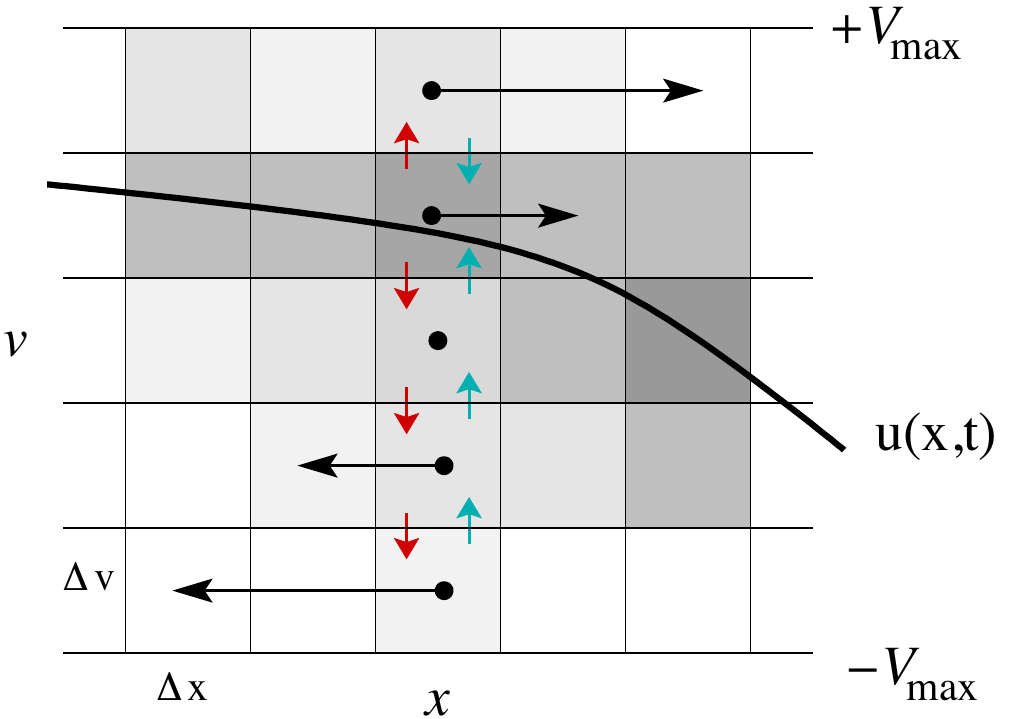}}
  \caption{Sketch of the lattice dynamics in the $(x,v)$
    position-velocity phase space. The solid curve is the fluid
    velocity profile; the greyscale tiling represents the
    discretization of particles mass in phase space. The black
    horizontal arrows show advection, while the blue and red vertical
    arrows are forcing and diffusion, respectively.}
  \label{fig:sketch_lattice}
\end{figure}

The solutions $f(\bm x, \bm v, t)$ to the Liouville (or
Fokker--Planck) equation~\eqref{eq:conslaw} are defined in the full
position-velocity phase-space $\Omega\times\mathbb{R}^d$, where
$\Omega$ designates a $d$-dimensional bounded spatial domain. To
simulate numerically the dynamics, we divide the phase-space in
$(2\times d)$-dimensional hypercubes. We then approximate
$f(\bm x, \bm v, t)$ as a piecewise-constant scalar field on this
lattice. Positions are discretized on a uniform grid with spacing
$\dx$ in all directions.  In principle, $f$ has to be defined for
arbitrary large velocities. We however assume that relevant values of
$\bm v$ are restricted to a bounded interval
$[-V_{\rm max},V_{\rm max}]^d$ , where $V_{\rm max}$ has to be
specified from physical arguments based on the forces $\mathcal{F}$
applied on the particles. Velocities are assumed to take $N_v^d$
values, so that the grid spacing reads $\dv = 2\,V_{\rm max}/N_v$.
Figure~\ref{fig:sketch_lattice} illustrates the phase-space
discretization in the one-dimensional case with $N_v=5$. The various
cells in position-velocity contain a given mass of particles. All
these particles are assumed to have a position and velocity equal to
that at the center of the cell.

The three phase-space differential operators appearing in
Eq.~\eqref{eq:conslaw}, namely the advection, the particle forcing,
and the diffusion, are applied one after the other, following an
\emph{operator splitting} method \citep{leveque2002finite}.  For the
advection step, we use a technique inspired from the Lattice-Boltzmann
method \citep[see, e.g., ][]{Succi:2001vh}. The time stepping is
chosen, so that a discrete velocity exactly matches a shift in
positions by an integer number of gridpoints. Namely, we prescribe
$\dx=\dv\,\dt$. All the particle phase-space mass located in
$[-\dv/2,\dv/2]$ does not move; that in $[\dv/2,3\dv/2]$ is shifted by
one spatial gridpoint to the right and that in $[-3\dv/2,-\dv/2]$ to
the left, etc.  All the mass is displaced from one cell to another
according to its own discrete velocity value. This evolution is
sketched by black horizontal arrows in
Fig.\ref{fig:sketch_lattice}. This specific choice for the
time-stepping implies that the advection (in space) is treated exactly
for the discrete system.  The next steps consist in applying the force
acting on the particles and the diffusion.  The corresponding terms in
Eq.~\eqref{eq:conslaw} are conservation laws, which suggests using a
finite-volume approximation. The time evolutions due to forcing and
diffusion are performed successively.  In both cases, we use classical
schemes (see below), where zero-flux conditions are imposed on the
boundary of $[-V_{\rm max},V_{\rm max}]^d$.  The force is evaluated
using the values of $\bm v$ at the centers of the cells and
$\nabla_v f$ is approximated using finite differences.  These steps
are illustrated by the horizontal blue and red arrows in
Fig.~\ref{fig:sketch_lattice}.

A few comments on the convergence and stability of the proposed
method. Clearly, except for specific singular forcings, all the linear
differential operators involved in \eqref{eq:conslaw} are expected to
be bounded.\footnote{Notice that, although the velocity might
  explicitly appear in the force $\mathcal{F}$, we only solve for a
  compact domain of velocities, thus preventing divergences.} We can
thus invoke the equivalence (or Lax--Richtmyer) theorem for linear
differential equations that ensures convergence, provided the scheme
is stable and consistent \citep[see][]{leveque2002finite}.

For the operator associated to particle acceleration, we use in this
study either a first-order upwind finite-volume scheme or a
higher-order flux limiter by following the strategy proposed
by~\citet{hundsdorfer1995positive}. The upwind scheme is first-order
accurate and is well-known for being consistent and stable if it
satisfies the Courant--Friedrichs--Lewy (CFL) condition.  This
requires that the time needed to accelerate particles by the grid size
$\dv$ is larger than the time step $\dt$. This leads to the stability
condition
\begin{equation}
  \mathrm{CFL} = (\dt/\dv)\,\max_{\bm x,\bm v,t}|\mathcal{F}(\bm x,\bm
  v,t)|/m_\mathrm{p} < 1.
  \label{eq:cfl}
\end{equation}
The upwind scheme is however known to suffer from numerical diffusion,
and obviously, one should only expect to recover the correct dynamics
only when the numerical diffusion $\kappa_{\rm num}$ is much smaller
than the physical one $\kappa$. The average numerical diffusion can be
estimated as
$\avg{\kappa_{\rm num}} \approx \avg{\mathcal{F}}\dt/\dv$. To limit
the effects of this numerical diffusion, we have also used a
flux-limiter scheme. While taking benefit of a higher-order
approximation where the field is smooth, it uses the ratio between
consecutive flux gradients to reduce the order in the presence of
strong gradients only.  The limiter is a nonlinear function of the
phase-space density field and the stability is ensured provided that
it is total-variation diminishing \citep[TVD;
see][]{leveque2002finite}. Among the various TVD limiters available on
the market, we choose the scheme of \citet{koren1993robust} with
parameter $2/3$.

For the term associated to diffusion, the flux at the interface
between two velocity cells is computed using finite differences. The
resulting finite-volume scheme is thus equivalent to compute a
discrete Laplacian on the velocity mesh. The stability condition is
then given by
\begin{equation}
\label{eq:stabLapl}
\frac{\kappa \dt}{\dv^2} < \frac{1}{2}.
\end{equation}
To summarize, the stability and convergence of the proposed method is
ensured when both \eqref{eq:cfl} and \eqref{eq:stabLapl} are
satisfied.

\medskip
From now on we restrict ourselves to small and heavy
particles whose interaction with the carrier fluid is dominated by
viscous drag and diffusion. In that case, we have
\begin{equation}
  \frac{\mathrm{d}\bm v_\mathrm{p}}{\mathrm{d}t} =
  -\frac{1}{\taup}\left(\bm v_\mathrm{p}-\bm u(\bm x_\mathrm{p},t)\right)
  +\sqrt{2\kappa}\,\bm \eta(t),
\label{eq:stokes_drag}
\end{equation}
where $\bm \eta(t)$ is the standard $d$-dimensional white noise and
the fluid velocity field $\bm u(\bm x,t)$ is prescribed and assumed to
be in a (statistically) stationary state.  This Stokes drag involves
the viscous particle response time
$\taup = 2 \rho_\mathrm{p} a^2/(9 \rho_\mathrm{f} \nu)$, with $a$ the
diameter of the particles, $\nu$ the viscosity of the fluid,
$\rho_\mathrm{p}\gg \rho_\mathrm{f}$ the particle and fluid mass
densities, respectively. Inertia is quantified by the Stokes number
$\St=\taup/\tau_{\rm f}$, where $\tau_{\rm f}$ is a characteristic
time of the carrier flow. The diffusion results from the random
collisions between the considered macroscopic particle and the
molecules of the underlying gas. Assuming thermodynamic equilibrium,
the diffusion coefficient reads
$\kappa = 2\,k_{\rm B}\,T/(m_\mathrm{p}\,\tau_\mathrm{p})$, where
$k_{\rm B}$ is the Boltzmann constant and $T$ the absolute
temperature.  The effect of diffusion is measured by the
non-dimensional number
$\mathcal{K} = \kappa\,\tau_{\rm f}/U_{\rm f}^2$ ($U_{\rm f}$ being a
characteristic velocity of the fluid flow).

Such a specific dynamics leads to appropriate estimates for the bound
$V_\mathrm{max}$ in particle velocity.  One can indeed easily check
that when $\kappa=0$, we always have
$|\bm v_\mathrm{p}| \le \max_{\bm x, t} |\bm u(\bm x,t)|$. In a
deterministic fluid flow, as for instance when $\bm u$ is stationary,
this gives the natural choice
$V_\mathrm{max} = \max_{\bm x, t} |\bm u(\bm x,t)|$. However, in most
situations, the maximal fluid velocity is not known a priori. One then
relies on the statistical properties of $\bm u$, as for instance its
root-mean square value $u_{\rm rms} = \langle u_i^2 \rangle^{1/2}$.
Usually the one-time, one-point statistics of fluctuating velocity
fields (being random or turbulent) are well described by a Gaussian
distribution. This ensures that by choosing
$V_\mathrm{max} = 3\,u_{\rm rms}$, the probability that a particle has
a velocity out of the prescribed bounds is less than 1\%.  Such
estimates are rather rough. In practice, it is known that the typical
particle velocity decreases as a function of the Stokes number. It was
for instance shown by \citet{abrahamson1975collision} that
$\langle |\bm v_\mathrm{p}|^2\rangle \propto u^2_{\rm rms} / \St$ at
very large Stokes numbers.  An efficient choice for $V_\mathrm{max}$
should account for that.

In the next two sections we investigate two different cases: First a
one-dimensional random Gaussian carrier flow with a prescribed
correlation time and, second, a two-dimensional turbulent carrier flow
that is a solution to the forced incompressible Navier-Stokes
equation.

%% SECTION  1D%%

\section{Application to a one-dimensional random flow}
\label{sec:1D}

%% SUBSECTION %%

\subsection{Particle dynamics for $d=1$}
\label{sec:1D_dyn}

In this section, we test our method in a one-dimensional
situation. For that, we assume that the fluid velocity is a Gaussian
random field, which consists in the superposition of two modes whose
amplitudes are Ornstein--Uhlenbeck processes, namely
\begin{eqnarray}
  u(x, t) &=&  A_1(t) \cos(2\pi\,x/L) +  A_2(t) \sin(2\pi\,x/L) \label{eq:u1d}\\
  \frac{\mathrm{d} A_i(t)}{\mathrm{d}t} &= &-\frac{1}{\tau_\mathrm{f}}
                                             A_i(t) \,
                                             +\sqrt{\frac{2}{\tau_\mathrm{f}}}\,
                                             \xi_i(t)
                                             \label{eq:A1d}
\end{eqnarray} 
where the $\xi_i$'s are independent white noises with correlations
$\langle \xi_i(t)\,\xi_i(t')\rangle=u_\mathrm{rms}^2\,\delta(t-t')$.
This flow is by definition fully compressible (potential) and
spatially periodic with period $L$. It is characterized by its
amplitude $\langle (u(x,t))^2\rangle^{1/2} = u_\mathrm{rms}$ and its
correlation time $\tau_\mathrm{f}$, which are fixed parameters. We
focus on the case when the Kubo number
$\Ku = \tau_\mathrm{f} \,u_\mathrm{rms}/L$ is of the order of unity.

\begin{figure}[ht]
  \begin{center}
    \includegraphics[width=0.48\columnwidth]{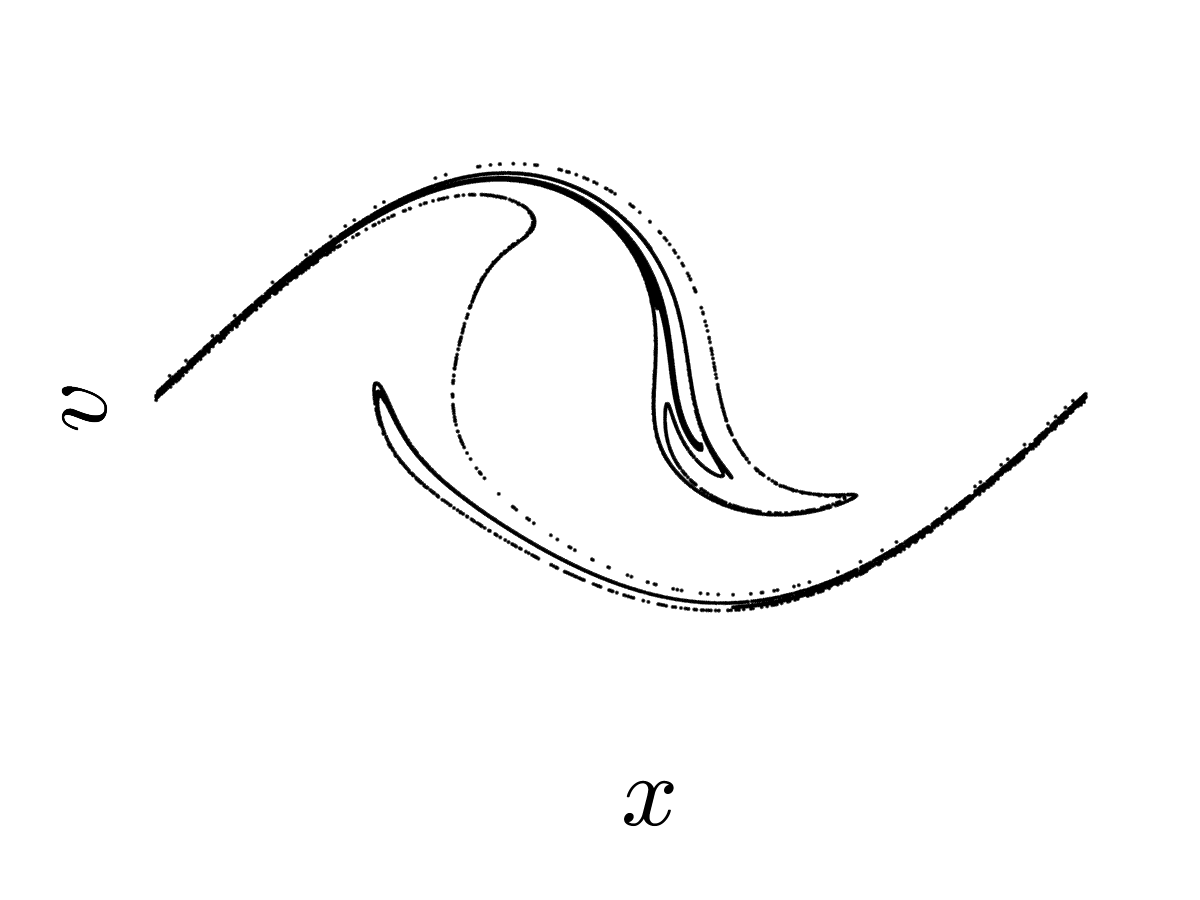}
    \includegraphics[width=0.48\columnwidth]{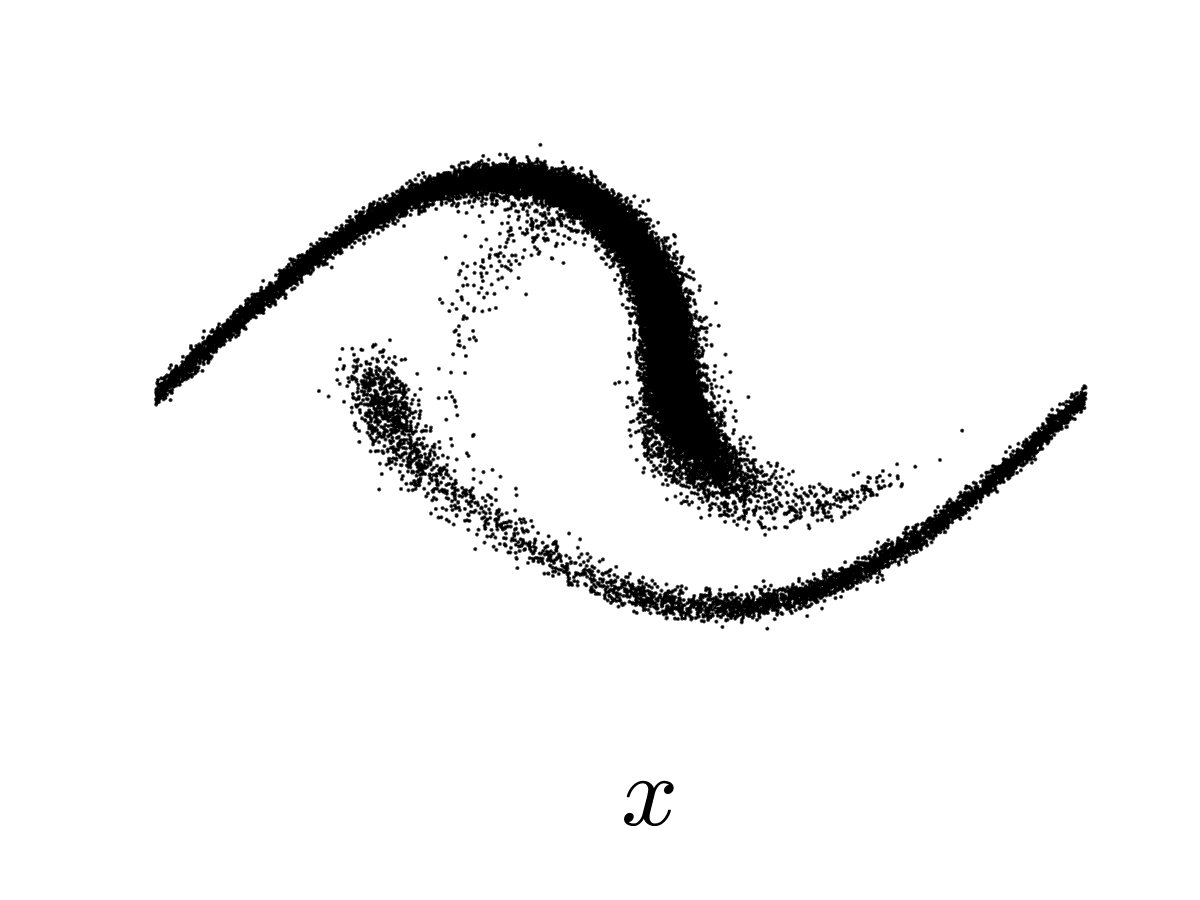}
  \end{center}
  \vspace{-25pt}
  \caption{Instantaneous snapshots of the particle positions in the
    $(x, v)$ plane for $\St\approx 2$ and with no diffusion
    ($\mathcal{K}=0$, Left) and with diffusion
    ($\mathcal{K} =6\times 10^{-4}$, Right). The folded structures are
    spread out by diffusion.}
  \label{fig:snapshot1d_diffkappa}
\end{figure}

\begin{figure}[ht]
  \centerline{\includegraphics[width=0.9\columnwidth]{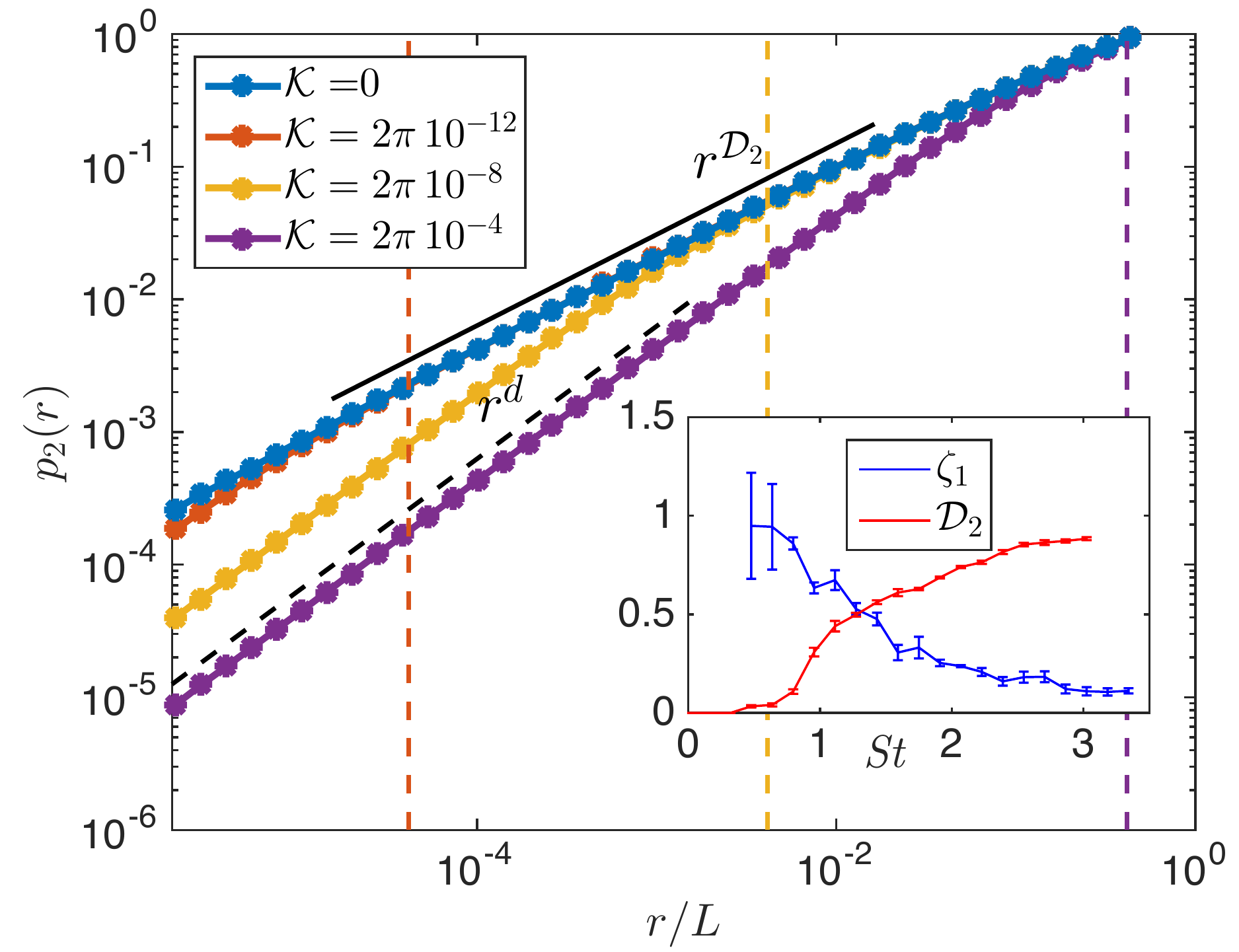}}
  \vspace{-5pt}
  \caption{ Cumulative probability $p_2^<(r)$ of inter-particle
    distances for various diffusivities $\kappa$ and for
    $\St \approx 2$.  One observes at low diffusivities and for
    $r>\ell_{\rm d}$ a behaviour $\propto (r/L)^{\D_2}$ with
    $\D_2 \approx 0.7 < d =1$, followed at small scales by uniform
    particle distribution where $p_2^<(r)\propto r$. As $\kappa$
    increases, the transition is moved to larger values of $r$. The
    vertical dashed represent in each case the estimate
    $\ell_d \sim \taup^{3/2}\kappa^{1/2}$ for this transition. Inset:
    variations of the correlation dimension $\D_2$ and of the scaling
    exponent $\zeta_1$ of the first order particle structure function,
    as a function of the Stokes number $\St$ in the case of the random
    fluid velocity defined by Eqs.~\eqref{eq:u1d}-\eqref{eq:A1d}.}
  \label{fig:1D_M2}
\end{figure}

We next consider particles suspended in this flow and following the
dynamics \eqref{eq:stokes_drag}.  The relevant Stokes number is then
defined as $\St = \tau_\mathrm{p}\,u_\mathrm{rms}/L$ and the relative
impact of diffusion is measured by
$\mathcal{K} = \kappa L/u_\mathrm{rms}^3$.  When diffusion is
neglected ($\mathcal{K}\to0$), the particles distribute on a dynamical
attractor (see Fig.\ref{fig:snapshot1d_diffkappa} Left) whose
properties depend strongly on $\St$.
These strange attractors are typically fractal objects in the phase
space and they are characterized by their fractal dimension spectrum
\citep{hentschel1983infinite}. The locations of particles are obtained
by projecting theses sets on the position space and might thus inherit
the associated clustering \citep{Bec:2003gf}.  The dimension that is
relevant for binary interactions between particles is the correlation
dimension $\D_2$, which relates to the probability of having two
particles within a given distance, namely
\begin{equation}
  p_2^{<}(r)=\mathbb{P}(|\bm x_\mathrm{p}^{(1)}(t)-\bm
  x_\mathrm{p}^{(2)}(t)|<r)\sim (r/L)^{\D_2},\text{ for } r\ll L,
  \label{eq:defD2}
\end{equation}
where $\bm x_\mathrm{p}^{(1)}$ and $\bm x_\mathrm{p}^{(2)}$ denote the
positions of two different particles.  Note that as we consider $u$ to
be in a statistically stationary state, $p_2^{<}$ is independent of
time . The correlation dimension $\D_2$ varies from $\D_2=0$ for a
point concentrations to $\D_2=1$ for a homogenous mass distribution .
In the example of Fig.~\ref{fig:snapshot1d_diffkappa} (Left)
$\D_2 \approx 0.7$. The variations of $\D_2$ as a function of the
Stokes number are displayed in the inset of
Fig.~\ref{fig:1D_M2}. $\D_2$ indeed varies from $0$ at small Stokes
numbers to values close to one.  For $\St=0$, the particles
concentrate on a point; their distribution is said to be atomic and
$\D_2 = 0$.  This is a consequence of the compressibility of the
one-dimensional (potential) flow.  Actually this behavior persists for
finite Stokes numbers, up to a critical value $\St^\star$, as shown by
\citet{wilkinson2003path} in the case where
$\tau_\mathrm{f}\ll L/u_{\rm rms}$ (that is $\Ku\to0$). We observe
here $\St^\star\approx 0.6$. For $\St>\St^\star$, the dimension
increases and tends to a homogenous distribution ($\D_2 = 1$) at large
particle inertia.

When one has only access to the Eulerian density of particles, the
distribution of distances cannot be directly inferred from
\eqref{eq:defD2}. One then relies on the coarse-grained density of
particles
\begin{equation}
  \rho_r(x,t) = \int_{-r/2}^{r/2} \mathrm{d}x' \int \mathrm{d}v \,
  f(x+x',v,t).
\label{eq:def_coarse}
\end{equation}
It is known that, under some assumptions on the ergodicity of the
particle dynamics, the second-order moment of this quantity scales as
$\langle \rho_r^2 \rangle \propto r^{\D_2-1}$ \citep[see,
e.g.,][]{hentschel1983infinite}. In one dimension, this second-order
moment is exactly the same as the radial distribution function. This
quantity will be used in the next sections to address the physical
relevance of the lattice-particle method.  It is of particular
interest when considering collisions between particles. Indeed, as
explained for instance by \citet{sundaram1997collision}, the
ghost-collision approximation leads to write the collision rate
between particles as the product of two contributions: one coming from
clustering and entailed in the radial distribution function, and
another related to the typical velocity differences between particles
at a given distance. This second quantity relates to the particle
velocity (first-order) structure function
\begin{equation}
  \Sone(r) = \left\langle| \bm v_\mathrm{p}^{(1)} - \bm v_\mathrm{p}^{(2)} |
    \,\middle | \, |\bm x_\mathrm{p}^{(1)} - \bm x_\mathrm{p}^{(2)}| = r
  \right\rangle.
  \label{eq:def_lagS1}
\end{equation}
This is the average of the amplitude of the velocity difference
between two particles that are at a given distance $r$. As the
probability of distances, this quantity behaves as a power law
$\Sone(r) \sim r^{\zeta_1}$ for $r\ll L$ \citep[see,
e.g.,][]{bec2005clustering}. The exponent $\zeta_1$, shown in the
inset of Fig.~\ref{fig:1D_M2} decreases from 1 at $\St=0$,
corresponding to a differentiable particle velocity field, to 0 when
$\St\to\infty$, which indicates that particle velocity differences
become uncorrelated with their distances. Again, when working with the
phase-space density one cannot use \eqref{eq:def_lagS1} but relies on
\begin{equation}
  \Sone(r) = \frac{\left\langle  \int \mathrm{d}v \int \mathrm{d} v'\,
      f(x, v)\, f(x+r, v')\, | v- v'| \right\rangle}{ \left\langle \int
      \mathrm{d}v \int \mathrm{d} v'\,
      f(x, v)\, f(x+r, v')\, \right\rangle}.
  \label{eq:def_eulS1}
\end{equation}
As the second-order moment of the coarse-grained density, this
quantity will also be used as a physical observable for benchmarking
the method.

In the above discussion, we have neglected the effects of
diffusion. It is for instance expected to alter clustering properties
by blurring the particle distribution at small scales.  This is
illustrated in Fig.~\ref{fig:snapshot1d_diffkappa} where one can
compare the instantaneous phase-space particle positions in the
absence of diffusion (Left) and when it is present (Right) at the same
time and for the same realization of the fluid velocity.  At large
scales, identical patterns are present, but diffusion acts at small
scale and smoothes out the fine fractal structure of the
distribution. One can easily estimate the scales at which this
crossover occurs. Diffusion is responsible for a dispersion
$v_{\rm d}$ in velocities that can be obtained by balancing Stokes
drag and diffusion in the particle dynamics, namely
$v_{\rm d}^2/\taup \approx \kappa $, so that
$v_{\rm d}\sim \taup^{1/2}\kappa^{1/2}$.  This dispersion in velocity
is responsible for a dispersion in positions on scales of the order of
$\ell_d = \taup\,v_{\rm d}\sim \taup^{3/2}\kappa^{1/2} =
\St^{3/2}\mathcal{K}^{1/2}L$.
Hence, when diffusion is small enough and $\ell_{\rm d}\ll L$, the
spatial distribution of particles is unchanged by diffusion at length
scales $r\gg\ell_{\rm d}$, and the probability that two particles are
at a distance less than $r$ behaves as $p_2^<(r) \sim (r/L)^{\D_2}$.
For $r\ll\ell_{\rm d}$, diffusion becomes dominant, the particles
distribute in a homogeneous manner and $p_2^<(r) \propto r^d$, with
$d=1$ being the space dimension. By continuity at $r=\ell_{\rm d}$, we
get $p_2^<(r) \sim (\ell_{\rm d}/L)^{\D_2}(r/\ell_{\rm d})$ at small
scales.

This picture is confirmed numerically as shown in Fig.~\ref{fig:1D_M2}
which represents the scale-behavior of $p_2^{<}(r)$ for a fixed Stokes
number and various values of the diffusivity $\kappa$.  One clearly
observes the homogenous distribution $\propto r^d$ at small scales and
the fractal scaling $\propto r^{\D_2}$ in an intermediate range. The
predicted transition between the two behaviors is indicated by the
vertical lines at the diffusive scale $\ell_{\rm d}$.  A homogeneous
distribution is recovered for $r\lesssim \ell_{\rm d}/10$.  

Velocity statistics are also altered by the presence of diffusion.
The structure function $\Sone(r)$ is expected to behave as $r^{\zeta_1}$
for $\ell_{\rm d}\ll r \ll L$ and to saturate to a constant value when
$r\ll\ell_{\rm d}$. By continuity, the value of this plateau should be
$\sim \ell_{\rm d}^{\zeta_1} \sim \mathcal{K}^{\zeta_1/2}$.  Note
finally that the slow convergence
$\ell_{\rm d}/L \propto \sqrt{\mathcal{K}}$ as $\mathcal{K}\to 0$
implies that very small values of the diffusion are needed in order to
clearly recover the statistics of diffusive-less particles as an
intermediate asymptotics.

%% SUBSECTION %%

\subsection{Lattice-particle simulations}
\label{sec:numerical setup}
		
We now turn to the application of the lattice-particle method
described in Sec.~\ref{sec:AlgDescr} to this one-dimensional
situation. We compare the results to Lagrangian simulations where we
track the time evolution of $N_p$ particles randomly seeded in space
with zero initial velocity.  We choose and normalize the initial
phase-space density $f(x,v,0)$ to match the Lagrangian settings. The
distribution is uniform over the cells, concentrated on a vanishing
velocity and the total mass is such that
$\sum_{i,j} f(x_i, v_j, t) \,\dx \,\dv = N_p$. In all simulations, the
maximum velocity is set to $V_{\rm max}=u_{\rm rms} =1$ and we have
chosen $L=2\pi$ and $\tau_{\rm f} = 1$.  In these units, the time step
is kept fixed at $\dt=\dx/\dv = 2^{-6}\pi\approx 0.05$. The number of
discrete velocities is of the form $2^{n}+1$ and is varied between
$N_v=3$ to $129$. The number of spatial collocation points is then
given by $2^{n+6}$ and thus varies between $N_x=128$ to $8192$.  Note
that, because of the CFL condition \eqref{eq:cfl}, this choice
restricts the number of discrete velocities that can be used to
$N_v<1+128\,\St$.

\begin{figure}[ht]
  \begin{center}
    \includegraphics[width=0.8\columnwidth]{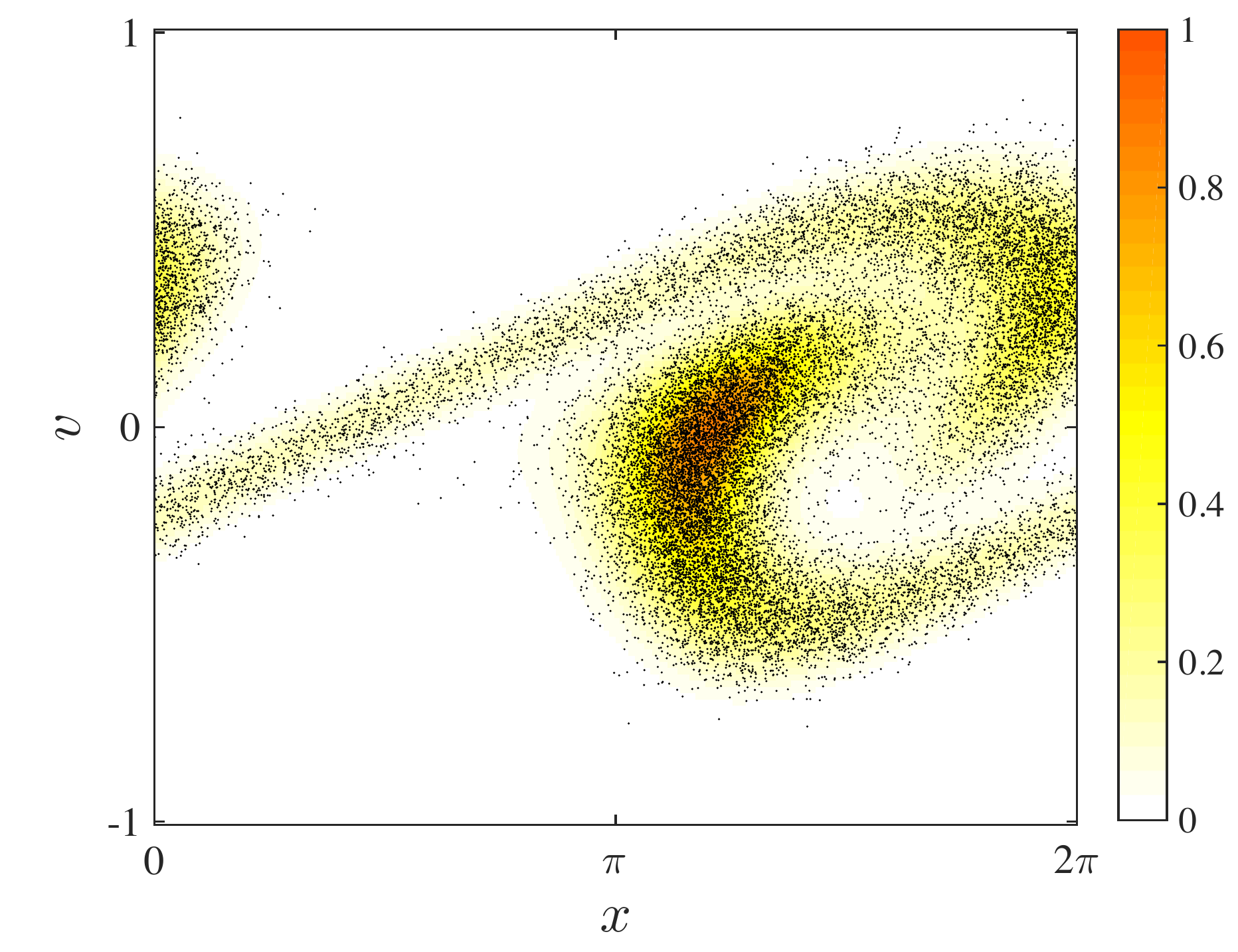}
  \end{center}
  \vspace{-15pt}
  \caption{Position-velocity phase-space positions of Lagrangian
    particles (black dots) on the top of the field obtained by the
    lattice-particle method (colored background). The diffusivity is
    here $\kappa=10^{-3}$ ($\mathcal{K} \approx 6.28.10^{-3}$) and
    $\St \approx 2$. }
  \label{fig:snapshot1d_with_without_diff}
\end{figure}
Figure \ref{fig:snapshot1d_with_without_diff} represents
simultaneously the phase-space distribution of Lagrangian particles
and the numerical approximation obtained by the particle-lattice
method for $N_v = 129$. Clearly, one observes that the model fairly
reproduces the distribution of particles, including the depleted
zones, as well as the more concentrated regions. Furthermore, the
method is able to catch multivalued particle velocities. We have for
instance up to three branches in $v$ for $x\simeq 3\pi/2$. It is
important to emphasize that numerical diffusion is of course present,
and that it has to be smaller than the physical diffusion $\kappa$ in
order for the model to be consistent with the Lagrangian dynamics.

To get a more quantitative insight on the convergence of the method,
we next compare the coarse-grained densities obtained from the
Lagrangian simulation and the lattice-particle approximation of the
phase-space density. The first, denoted $\rho_r^{\rm L}$ is computed
by counting the number of particles contained in the different boxes
of a tiling of size $r$. The second is written as $\rho_r^{\rm E}$ and
is obtained by summing over velocities and coarse-graining over a
scale $r$ the phase-space density obtained numerically. To confirm the
convergence of the method, we measure for a fixed $r$ the behavior of
the $\Ld$-norm of the difference between $\rho_r^{\rm L}$ and
$\rho_r^{\rm E}$, namely
\begin{equation}
  \| \rho_r^{\rm L} - \rho_r^{\rm E} \| = \left \langle
    \left(\rho_r^{\rm L}(x,t) - \rho_r^{\rm E}(x,t) \right)^2 \right \rangle ^{1/2},
\end{equation}
where the angular brackets $\langle \cdot \rangle$ encompass a spatial
and a time average.
\begin{figure}[ht]
  \centerline{\includegraphics[width=0.8\columnwidth]{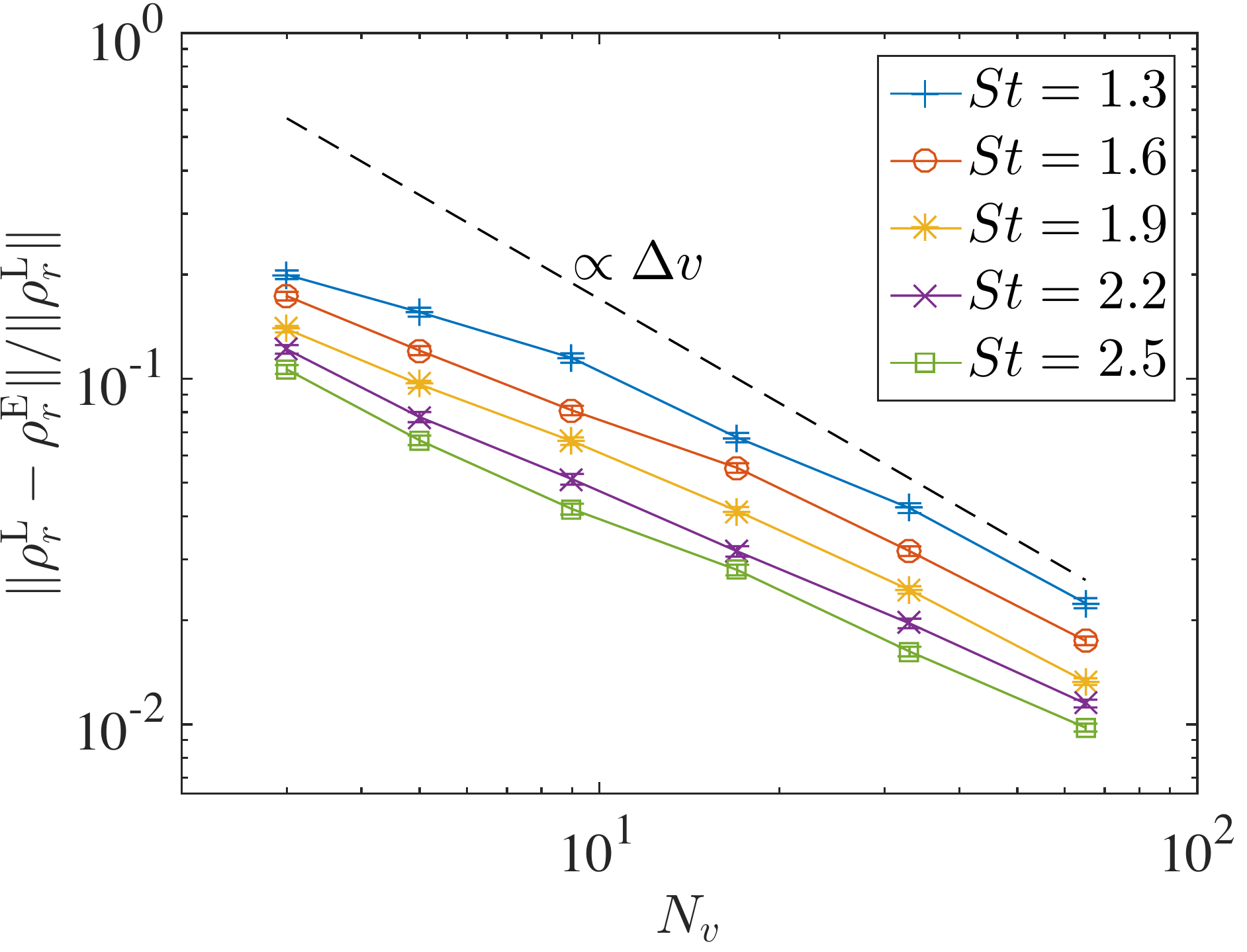}}
  \vspace{-10pt}
  \caption{ Relative $\mathcal{L}_2$-error of the lattice-particle
    method for evaluating the coarse-grained density $\rho_r$ over a
    scale $r = L /128$ as a function of the number of velocity
    gridpoints $N_v$ and for various values of the Stokes number, as
    labeled.}
  \label{fig:l2_convergence}
\end{figure}
Figure~\ref{fig:l2_convergence} shows the behavior of the relative
$\Ld$-error as a function of the number of velocity gridpoints $N_v$,
for various values of the Stokes number $\St$ and for a given scale
$r$. One observes that the error decreases when the resolution
increases, giving strong evidence of the convergence of the method.
The error is found proportional to the velocity grid spacing
$\Delta v$, indicating that the method is first order. The constant is
a decreasing function of the Stokes number. This indicates that the
method is more accurate for particles with strong inertia. The reason
for this trend will be addressed in the sequel.

To assess the ability of the proposed method to reproduce physically
relevant quantities, we now compare statistics obtained using the
lattice method with those using a Lagrangian approach. We focus on the
clustering and velocity difference properties that were introduced and
discussed in Section~\ref{sec:1D_dyn}.

\begin{figure}[ht]
  \centerline{\includegraphics[width=0.8\columnwidth]{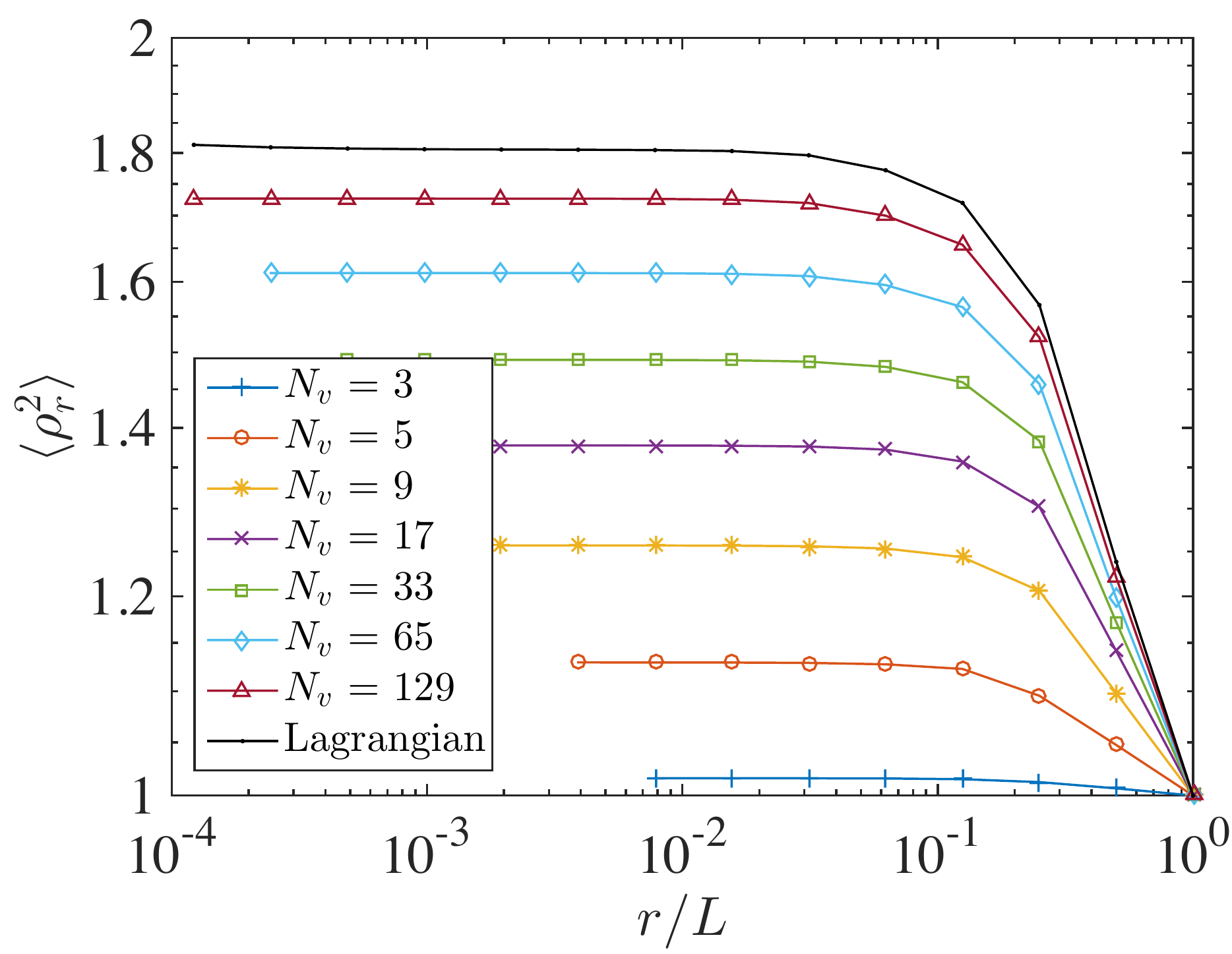}}
  \vspace{-10pt}
  \caption{ Convergence of the second-order moment of the
    coarse-grained density $\langle \rho_r^2\rangle$, which is shown
    as a function of $r$ for $\St \approx 1.9$,
    $\mathcal{K}=\pi\,10^{-2}$, and various lattice velocity
    resolutions $N_v$, as labeled.}
  \label{fig:1D_M2_convergence}
\end{figure}
Figure~\ref{fig:1D_M2_convergence} shows for given values of the
Stokes number and of the diffusivity, the second-order moment of the
coarse-grained density $\langle(\rho_r^{\rm E})^2\rangle$ as a
function of $r$ and various values of the resolution in velocities,
together with the value $\langle(\rho_r^{\rm L})^2\rangle$ obtained
with $10^6$ Lagrangian particles.  One observes that the curves
approach the limiting behavior from below when the number of
gridpoints $N_v$ becomes larger (i.e.\ when $\Delta v\to 0$). At
sufficiently high velocity resolutions, the method is able to capture
the large-scale properties of the concentration of the particles.  The
second-order moment of density then saturates to a value lower than
that expected from Lagrangian measurements. The situation is very
different at very low resolutions where the data obtained from the
lattice-particle method deviates much, even at large scales. This
corresponds to the case when the numerical diffusion in velocity is
larger than the physical diffusion.

These strong deviations stem from a non-trivial effect of diffusion
that lead to finite-scale divergences of the solutions associated to
different values of $\mathcal{K}$.  In the absence of diffusion, there
is a finite probability that an order-one fraction of mass gets
concentrated on an arbitrary small subdomain of the position-velocity
phase space. This corresponds to a violent fluctuation where the local
dimension approaches zero. At the time when this occurs, the mass
distribution associated to a finite value of the diffusion will get
stacked at a scale $\ell_\mathrm{d}$. Because of the chaotic nature of
the particle dynamics, the two mass distributions, with and without
diffusion will experience very different evolutions and diverge
exponentially fast. 
\begin{figure}[ht]
  \begin{center}
    \includegraphics[width=0.3\columnwidth, height=0.4\columnwidth]{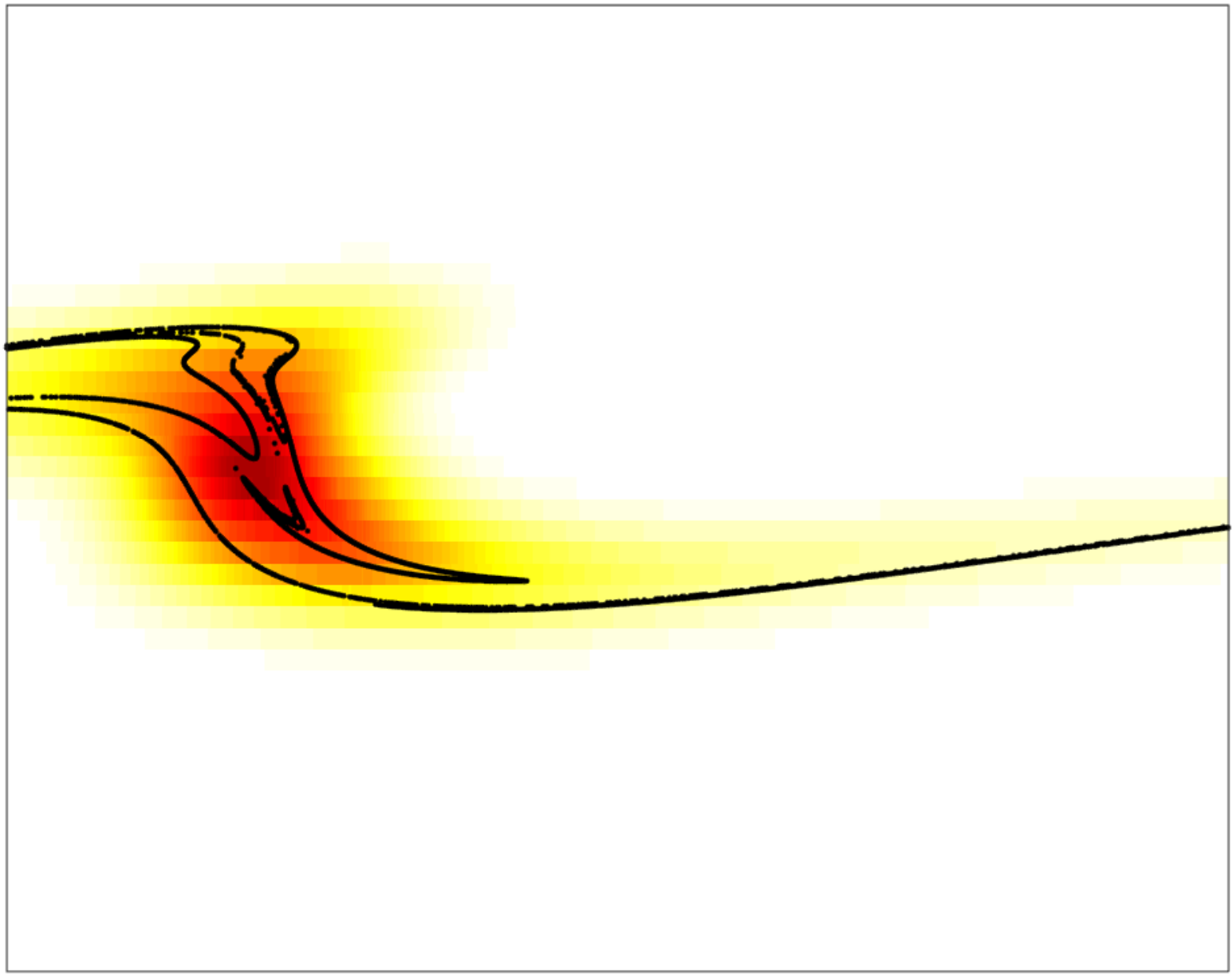}
    \includegraphics[width=0.3\columnwidth, height=0.4\columnwidth]{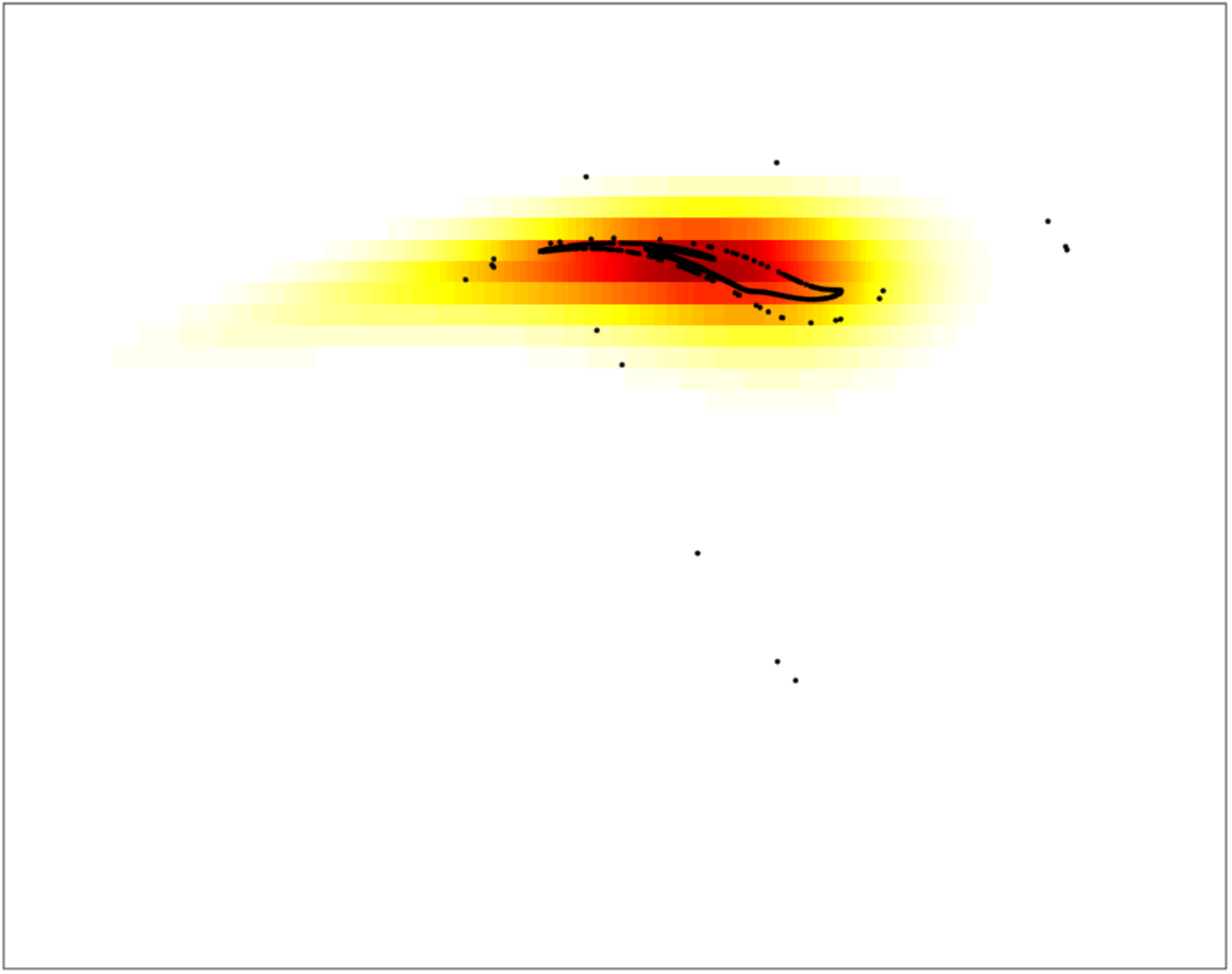}
    \includegraphics[width=0.3\columnwidth, height=0.4\columnwidth]{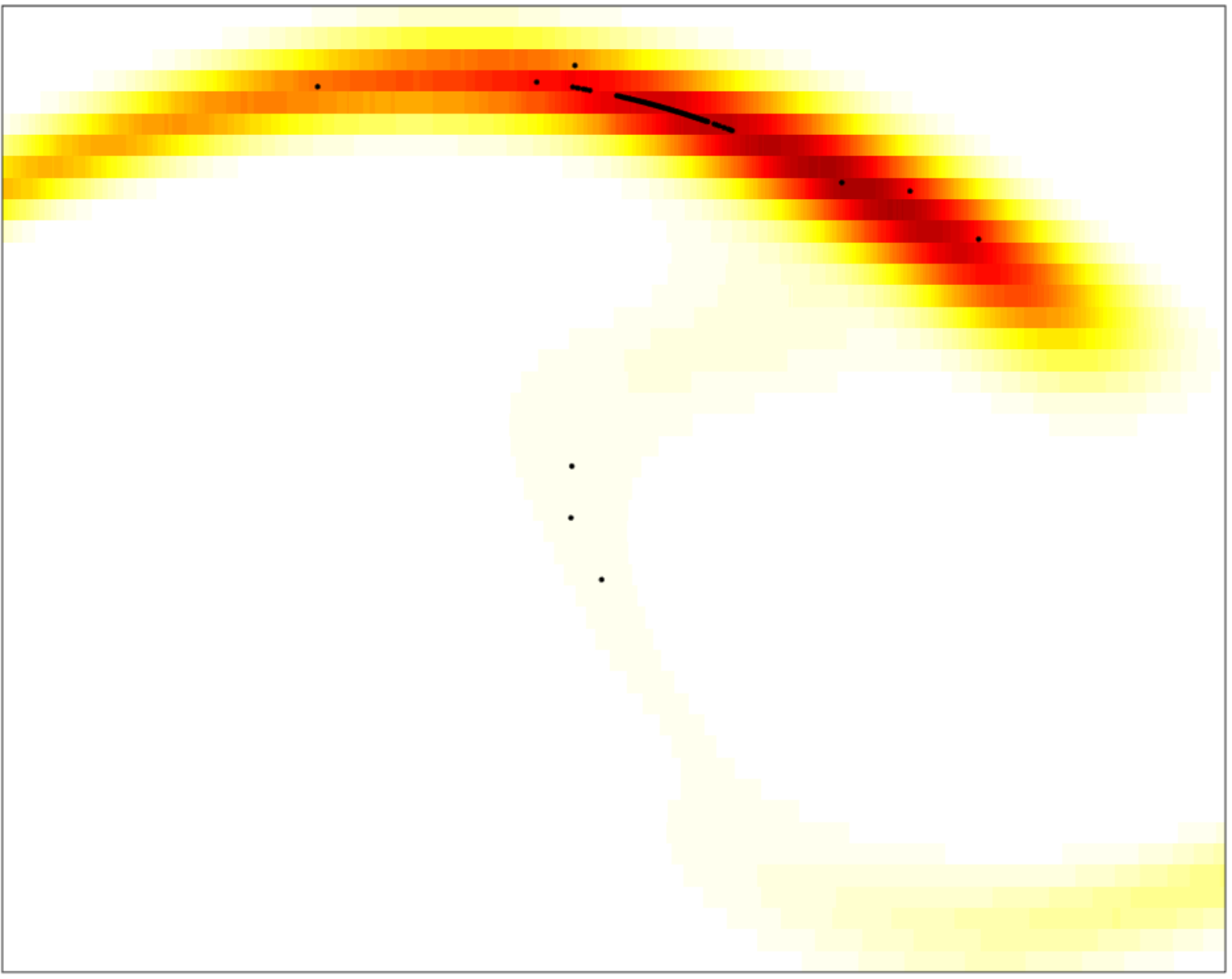}
  \end{center}
  \vspace{-15pt}
  \caption{Three snapshots of the Lagrangian particles (black dots,
    for $\mathcal{K}=0$) and of the lattice-particle Eulerian solution
    (colored background) in the $(x,v)$ plane for different times: At
    $t<t_\star$ (Left) the solution is well approximated at large
    scales; At $t=t_\star$ (Center) an order-unity fraction of the
    mass is concentrated on a scale less than $\ell_{\rm d}$; At
    $t>t_\star$ (Right), the Eulerian and Lagrangian solutions diverge
    exponentially fast with differences appearing at the largest
    scales..}
  \label{fig:massdiverge}
\end{figure}
Such a strong clustering event followed by the divergence of the
solutions, is shown in Fig.~\ref{fig:massdiverge}. Starting from a
correctly reproduced distribution, the major part of non-diffusive
Lagrangian particles concentrate into a subgrid region while the
Eulerian approximation is stacked at scales of the order of
$\ell_{\rm d}$.  At a later time, the two distributions diverge and
the diffusive particles fill faster larger scales. The probability
with which one encounters such a configuration strongly depends on the
Stokes number and on the spatial dimension. In the one-dimensional
case, such events are rather frequent but become sparser when the
Stokes number increases. This is essentially due to the
compressibility of the carrier flow. For incompressible fluids in
higher dimensions, we expect a negligible contribution from these
events.

\begin{figure}[ht]
  \centerline{\includegraphics[width=0.8\columnwidth]{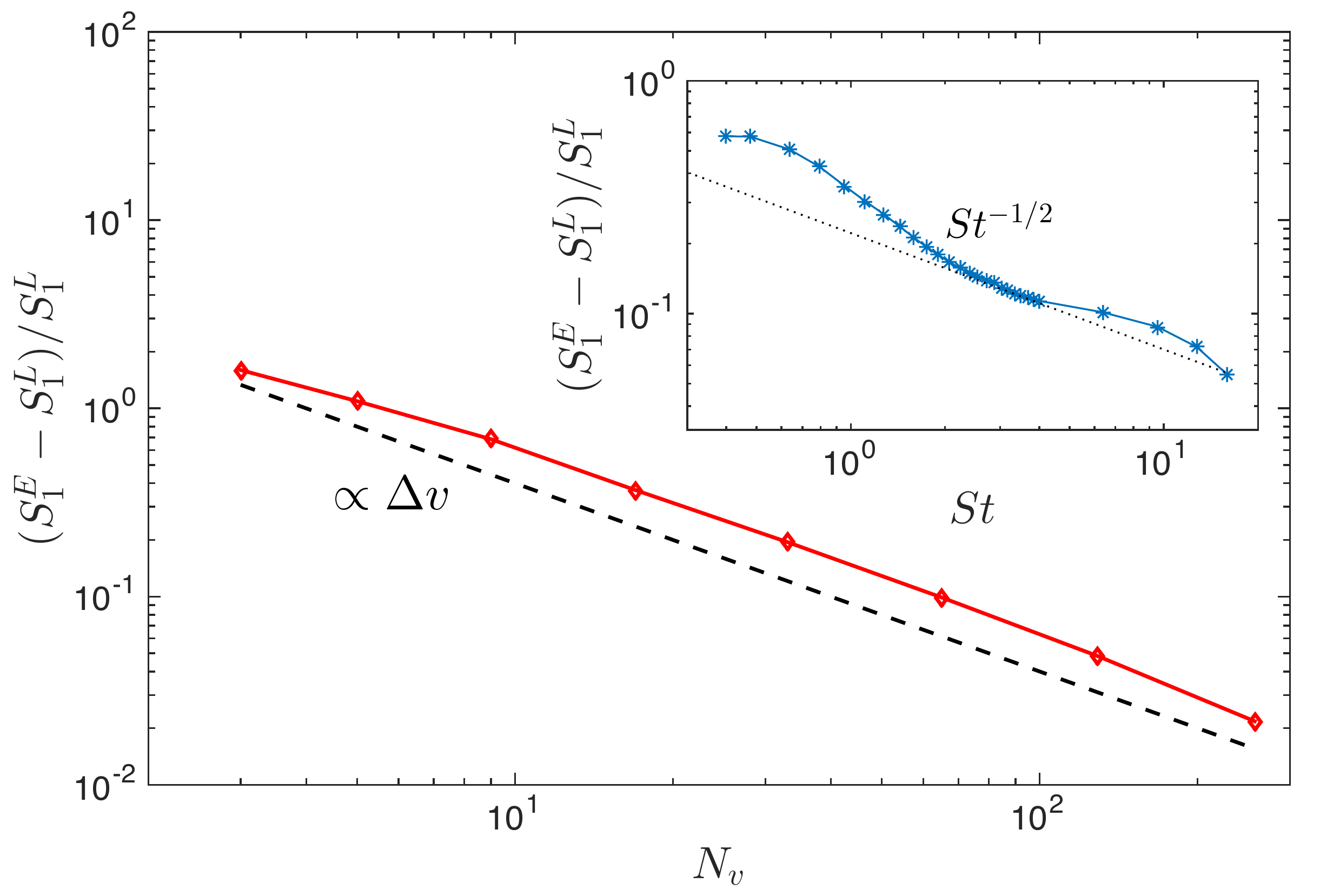}}
  \vspace{-10pt}
  \caption{Relative error between the particle velocity structure
    function $\Sone^E(r)$ obtained from the lattice-particle method and
    that $\Sone^L(r)$ from Lagrangian averages, as a function of the
    number $N_v$ of velocity gridpoints.  Here, the Stokes number is
    fixed $\St \approx 1.9$ and $\mathcal{K} = 2\pi\,10^{-3}$.  Inset:
    same quantity but for $N_v=33$ and as a function of the Stokes
    number $\St$.}
  \label{fig:1D_coll_convergence}
\end{figure}

To close this section on one-dimensional benchmarks of the
lattice-particle method, we report some results on velocity difference
statistics. For that, we have measured the first-order structure
function $\Sone(r)$ of the particle velocity, using
(\ref{eq:def_lagS1}) in the Lagrangian case and (\ref{eq:def_eulS1})
for solutions obtained with the lattice-particle
method. Figure~\ref{fig:1D_coll_convergence} shows the relative error
of $\Sone(r)$ for fixed values of the separation $r$, the Stokes
number, and the diffusivity, as a function of the velocity
resolution. Clearly, when the number of gridpoints $N_v$ increases,
the error decreases, following a law approximatively proportional to
the grid spacing $\Delta v$. The inset shows the same quantity but,
this time, for a fixed resolution ($N_v=33$) and as a function of the
Stokes number.  One clearly observes a trend for this error to
decrease with $\St$. There are two explanations for this
behavior. First, as seen above, there are strong clustering events
leading to differences between the Lagrangian and lattice solutions
that can persist for a finite time. When the Stokes number increases,
such events become less probable. The second explanation relies on the
fact that particles with a larger Stokes number experience weaker
velocity fluctuations. This implies that for a fixed value of
$V_{\rm max}$, the particle velocity is more likely to be fully
resolved at large values of $\St$.  As seen in the inset of
Fig.~\ref{fig:1D_coll_convergence}, the downtrend of the error is
compatible with a behavior $\propto \St^{-1/2}$. It might thus be
proportional to the expected value of the root-mean-squared particle
velocity when $\St\gg 1$ \citep{abrahamson1975collision}, favoring the
second explanation. We will turn back in Sec.~\ref{sec:direct} for the
two-dimensional case on the effect of the maximal resolved velocity
$V_{\rm max}$ onto the convergence of the lattice-particle method.

%% SECTION %%

\section{Application to incompressible two-dimensional flows}
\label{sec:2D}

We extend in this section the lattice-particle method to
two-dimensional flows.  For the particle acceleration, we again make
use of a flux-limiter scheme. 

				%% SUBSECTION %%
\subsection{Cellular flow}
We first consider a fluid flow that is a stationary solution to the
incompressible Euler equations (and to the forced Navier--Stokes
equations).  It consists of a cellular flow field, a model that have
often been used to investigate mixing properties, as well as the
settling of heavy inertial particles \citep[see,
e.g.,][]{maxey1986gravitational,bergougnoux2014motion}. The velocity
field is the orthogonal gradient of the $L$-periodic bimodal stream
function $\psi(x,y) = U\,\sin(\pi(x+y)/L)\sin(\pi(x-y)/L)$ (the
typical velocity strength is here denoted by $U$). The cellular flow
has been here tilted by an angle $\pi/4$ in order to avoid any
alignment of the separatrices between cells with the lattice that
leads to spurious anisotropic effects.
	
\begin{figure}
  \includegraphics[width=0.23\textwidth,height=0.23\textwidth]{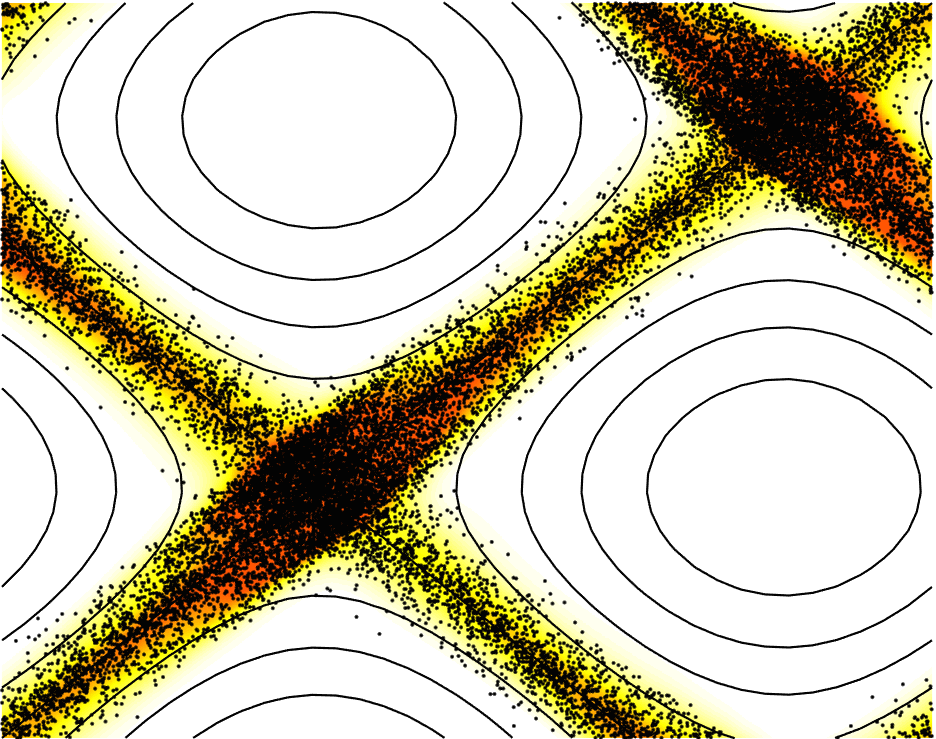}
  \
  \includegraphics[width=0.23\textwidth,height=0.23\textwidth]{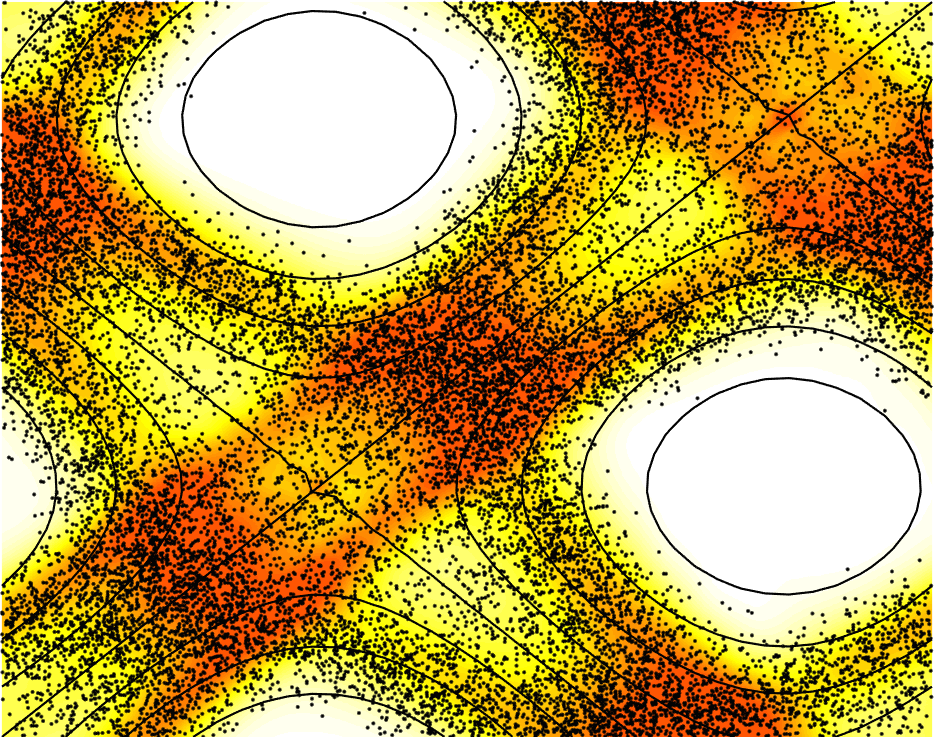}
  \caption{Particles stationary distribution inside a tilted cellular
    flow along with the density field from the lattice method. The
    value of the diffusivity is $\mathcal{K}=8\pi\times10^{-3}$. Left:
    $\St = 1/(2\pi)$. Right: $\St = 1/\pi$. These simulations were
    performed on a lattice with $1024^2$ spatial gridpoints associated
    to $19^2$ discrete velocities. The distributions have been here
    spatially shifted in order to avoid having the concentration point
    $(0,0)$ at the origin.}
  \label{fig:2DSNAPCELL}
\end{figure}
Figure \ref{fig:2DSNAPCELL} shows two snapshots for two different
values of $\St = \taup U/L$ of the stationary particle distribution
(black dots), together with the density field evolved by the
lattice-particle dynamics. For the smallest Stokes number (Left
panel), one observes that the particle distribution is concentrated
along the separatrices between the different cells. One also observes
that it develops entangled structures in the vicinity of the
hyperbolic stagnation points of the flow. These loops, which are
aligned with the stable direction, corresponds to oscillations in the
particle dynamics that occurs when their inertia makes them cross the
unstable manifold with a too large velocity. At larger $\St$, the
particle distribution is somewhat broader but is this time centered on
specific trajectories that do not perform the aforementioned
oscillation but rather cross ballistically the heteroclinic
separatrices. These results show that the lattice-particle method is
able to reproduce the complex dynamics of particles in a
two-dimensional steady flow. The fine structures of the spatial
distribution are fairly reproduced, as long as the numerical diffusion
surpasses numerical errors.

\subsection{Heavy particles in 2D turbulence}

We next turn to the study of the model in non-stationary fluid flows
that are solutions to the forced two-dimensional incompressible
Navier--Stokes equations
\begin{equation}
  \label{eq:ns2D}
  \partial_t \bm u + \bm u\cdot\nabla\bm u = -\nabla p + \nu\nabla^2 \bm u -
  \alpha \bm u + \bm f, \quad \nabla\cdot\bm u = 0.
\end{equation}
The linear damping term involving the coefficient $\alpha$ originates
from Ekman friction (in geophysical flows), Rayleigh friction (in
stratified fluids) or the friction induced by the surrounding air in
soap-film experiment. The flow is maintained in a statistical steady
state by the forcing $\bm f$ that is assumed to be concentrated over a
specific scale $\ell_{\rm f}$. The fluid velocity field $\bm u$ is
computed numerically using a pseudo-spectral, fully de-aliased GPU
solver for the vorticity-streamfunction formulation of the
Navier-Stokes equation (\ref{eq:ns2D}). 

The two-dimensional Navier--Stokes equation is known to develop two
cascades \citep[see, e.g.,][]{boffetta2012two}. Kinetic energy
undergoes an \emph{inverse cascade} with a rate $\varepsilon$ toward
the large scales $\ell\gg\ell_{\rm f}$ where it is dissipated by the
linear damping. The enstrophy $\langle \omega^2\rangle$ (where
$\omega = \partial_x u_y-\partial_y u_x$ designates the vorticity)
experiences a \emph{direct cascade} to the small scales
$\ell\ll\ell_{\rm f}$ with a rate $\eta$ and is then dissipated by
molecular viscosity. These different cascades are associated to
different behaviors of the velocity power spectrum. For
$k\ll k_{\rm f} \propto \ell_{\rm f}^{-1}$, the inverse energy cascade
promotes a $k^{-5/3}$ Kolmogorov law, as in the three-dimensional
direct cascade. At small scales, i.e.\ for $k\gg k_{\rm f}$ in the
direct enstrophy cascade, the flow is characterized by long-living
vortices and the spectrum follows a $k^{-3}$ Batchelor--Kraichnan law
with a logarithmic correction.

\begin{figure}[ht]
  \centerline{\includegraphics[width=0.8\columnwidth]{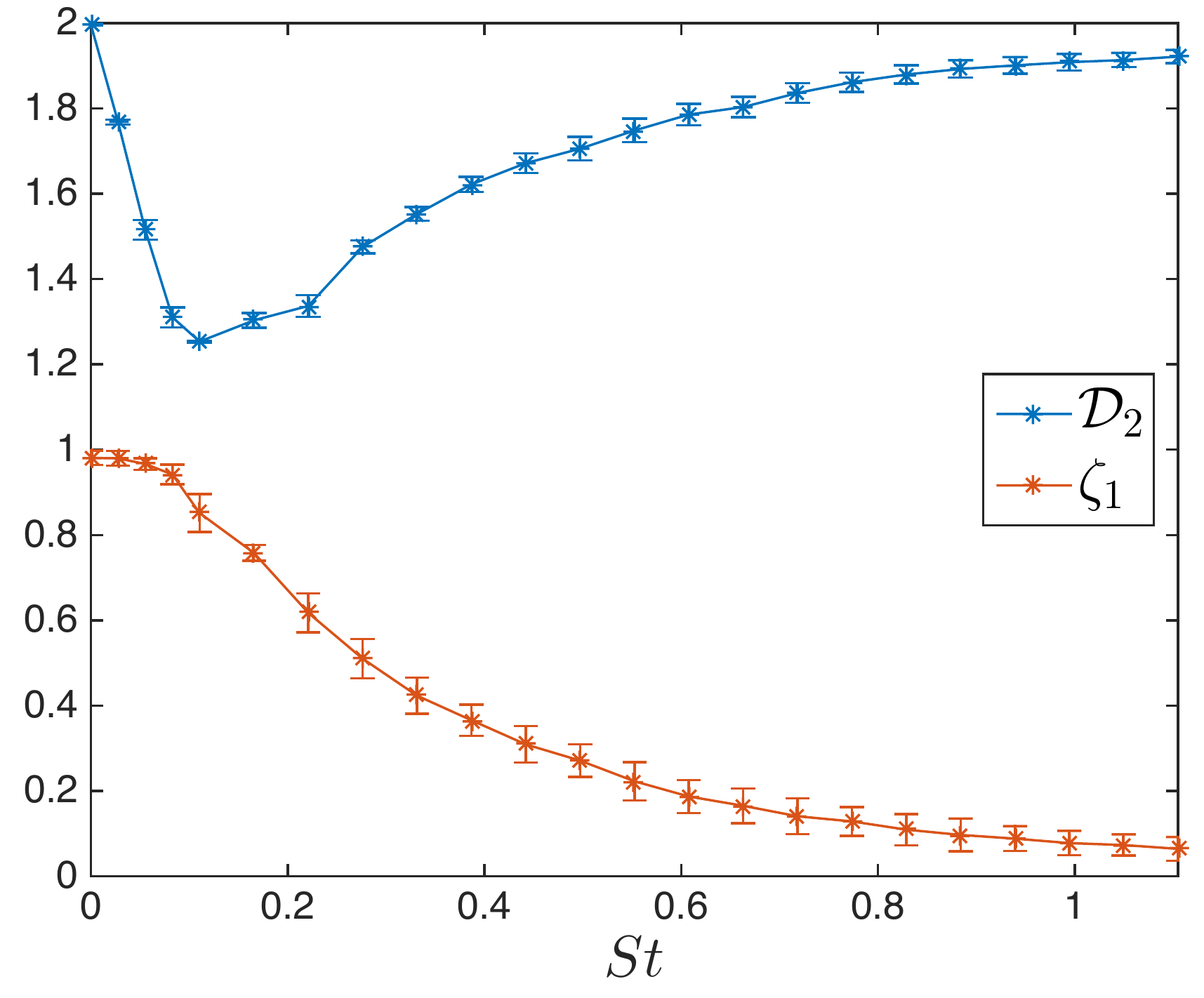}}
  \caption{Correlation dimension $\D_2$ and scaling exponent $\zeta_1$
    of the particle velocity first-order structure function as a
    function of $\St$ in the two-dimensional direct enstrophy
    cascade.}
  \label{fig:2D_D2}
\end{figure}

Dimensional analysis predicts that the direct enstrophy cascade is
associated to a unique timescale $\tau_{\Omega} = \eta^{-1/3}$.
Investigating heavy particle dynamics at the small scales of
two-dimensional turbulence thus requires comparing their response time
to $\tau_{\Omega}$. The relevant parameter is then the Stokes number
defined as $\St = \taup/\tau_\Omega$.  For $\St\ll 1$, particles
almost follow the flow and tend to distribute homogeneously in space.
When $\St\gg1$, they completely detach from the fluid and experience a
ballistic motion leading again to a space-filling
distribution. Non-trivial clustering effects occur when the Stokes
number is order one.
\begin{figure}[ht]
  \begin{center}
    \includegraphics[width=0.85\columnwidth, height=0.85\columnwidth]{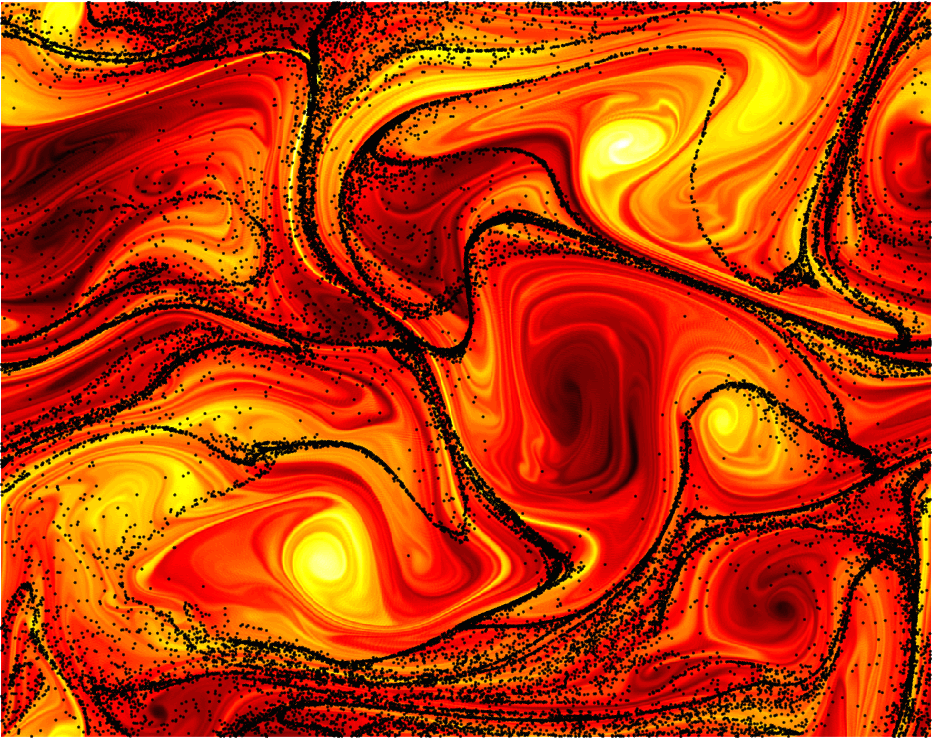}
  \end{center}
  \caption{Snapshot of the position of particles (black dots) for
    $\St = 0.1$. The colored background shows the vorticity
    field obtained from a $1024^2$ direct numerical simulation with a
    large-scale forcing at wavenumbers $1\le |\bm k|<4$.}
  \label{fig:snapshot2d}
\end{figure}
This is illustrated in Fig.~\ref{fig:snapshot2d}, which shows a
snapshot of the particle distribution in the position space on top of
the turbulent vorticity field in the direct enstrophy cascade. Due to
their inertia, particles are ejected from vortices and concentrate in
high-strain regions. There, the combination of stretching, folding and
dissipation induced by their dynamics makes them converge to a
dynamical attractor with fractal properties. Such a behavior is
quantitatively measured by the correlation dimension $\D_2$ defined in
Eq.~(\ref{eq:defD2}).  The evaluation of $\D_2$ as a function of $\St$
resulting from Lagrangian simulations is presented in
Fig.~\ref{fig:2D_D2}.  At $\St = 0$, unlike in the one-dimensional
case where the dimension of the attractor is $0$, particles follow the
streamlines of the incompressible two-dimensional flow, fill the
position space, and hence $\D_2=2$. Clustering then increases with
inertia to attain a minimum at $\St \approx 0.2$. It then decreases
again as the velocity of particles separate from that of the fluid and
disperse in the velocity space, leading to a space-filling
distribution $\D_2=2$ when $\St\to\infty$.

The velocity distribution of particles is itself having a behavior
that is very similar to the one-dimensional case. This is clear from
Fig.~\ref{fig:2D_D2}, where the scaling exponent $\zeta_1$ of its
first-order structure function (see Eq.~(\ref{eq:def_lagS1})) is
represented as a function of the Stokes number. For $\St\ll1$, the
particles are as if advected by a smooth velocity field and
$\zeta_1\approx1$. When $\St\gtrsim 1$, particles with very different
velocities can come arbitrarily close to each other and $\zeta_1\to0$.

Particle properties in the inverse energy cascade are more difficult
to characterize because of the scale-invariance of the velocity
field. In particular, neither the moments of the coarse-grained
density nor the particle velocity structure functions display any
scaling behavior. What has been nevertheless observed numerically by
\cite{boffetta2004large} is that the particle spatial distribution is
dominated by the presence of voids whose sizes obey a universal
scaling law. \cite{chen2006turbulent} argued that such voids are
related to the excited regions of the flow and that particles tend to
follow the calm regions where the zeros of the fluid acceleration are
more probable.

In the sequel we apply the lattice method to both the direct and the
inverse two-dimensional cascades. Resolving both cascades in the same
simulation would require a tremendous scale separation and thus number
of gridpoints \citep[see][]{boffetta2010evidence}. For that reason we
consider the two cases separately.

\subsubsection{Direct enstrophy cascade}
\label{sec:direct}

\begin{figure}[ht]
  \begin{center}
    \includegraphics[width=0.85\columnwidth, height=0.85\columnwidth]{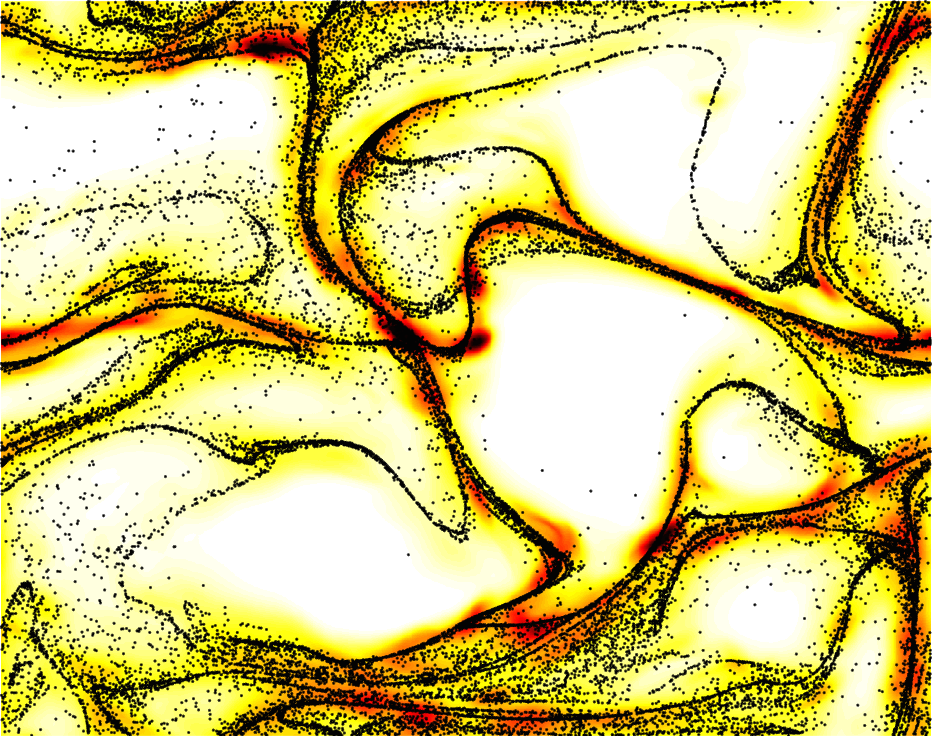}
  \end{center}
  \caption{Snapshot of the position of Lagrangian particles (black
    dots), together with the density field obtained from the lattice
    method (colored background, from white: low densities to red: high
    densities) for $\St\approx 0.1$ at the same instant of time as
    Fig.~\ref{fig:snapshot2d}. The fluid flow was integrated using a
    resolution of $1024^2$ and lattice simulations were performed with
    $1024^2$ spatial gridpoints associated to $17^2$ discrete
    velocities.}
  \label{fig:2D_LAG_LAT}
\end{figure}

The fluid flow is integrated by a pseudo-spectral method on a uniform
square spatial grid using a stream-function formulation of the
Navier-Stokes equation~(\ref{eq:ns2D}).  To maintain a developed
turbulent state, a stochastic forcing is applied in the wavenumber
shell $1\le|\bm k|<4$ of Fourier space while the kinetic energy
accumulating at large scales is removed by a linear friction. The
particle dynamics is simulated using a spatial lattice with the same
resolution as the fluid and with various numbers $N_v^2$ of discrete
velocities. The acceleration step is done via a flux limiter scheme as
described in Sec.~\ref{sec:AlgDescr}. Results are compared to particle
trajectories obtained from Lagrangian
simulations. Figure~\ref{fig:2D_LAG_LAT} shows the instantaneous
particle distributions obtained from the two approaches. The
qualitative agreement is excellent, reproducing correctly depleted
zone as well as more concentrated regions.

\begin{figure}[ht]
  \centerline{\includegraphics[width=0.8\columnwidth]{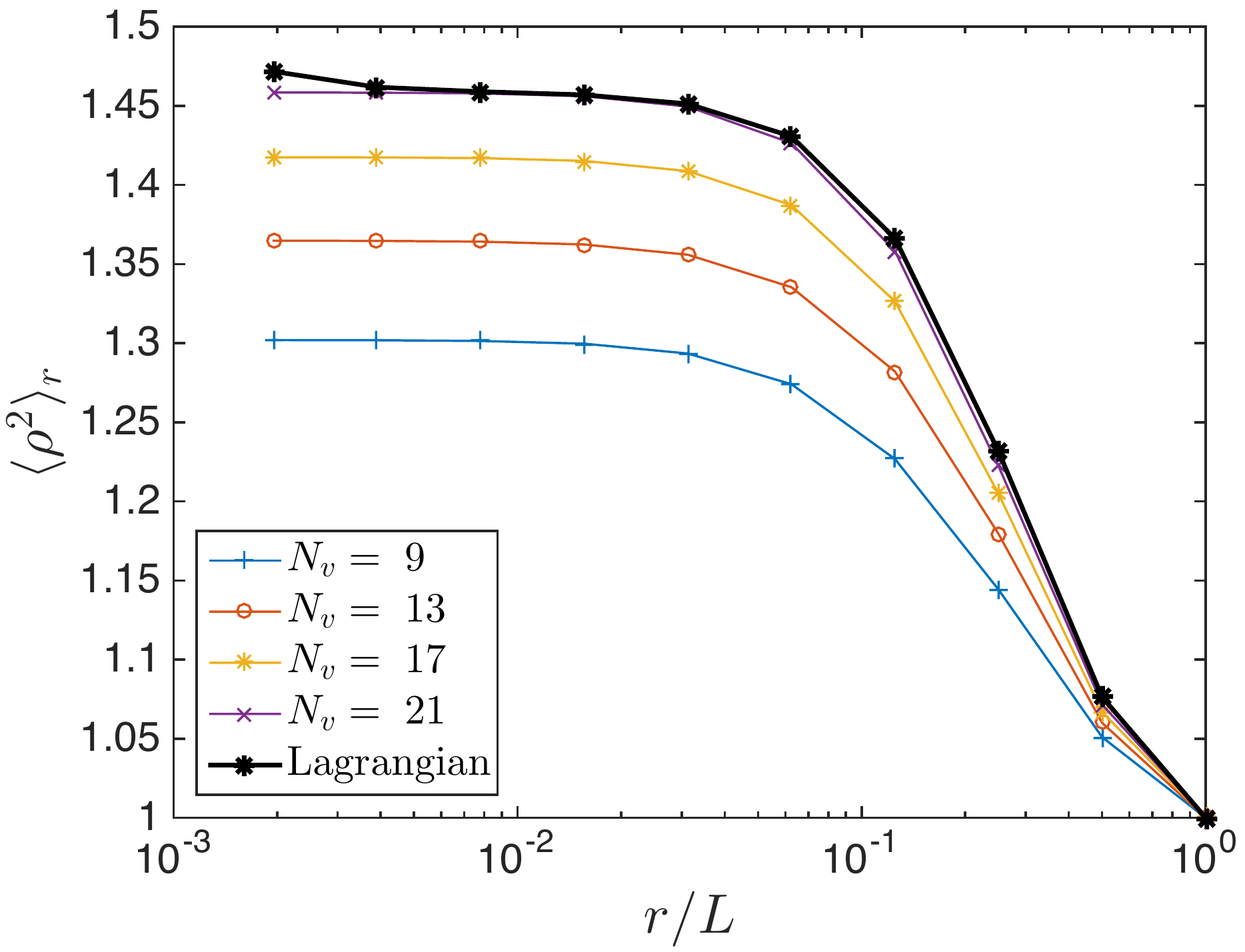}}
  \caption{Second-order moment of the coarse-grained density $\rho_r$
    as a function of $r$ for $\St=0.5$ and $\mathcal{K} = 1.6 \pi \; 10^{-1} $
    in the direct cacade, both from Lagrangian measurement (black
    line) and the lattice method with different $N_v$, as labeled.}
  \label{fig:2D_meM2}
\end{figure}

To get more quantitative informations on the relevance of the method,
we have performed a set of simulations with a $512^2$ resolution and
in which both the number of discrete velocities $N_v^2$ and the maximum
velocity $V_{\rm max}$ are varied. Figure~\ref{fig:2D_meM2} shows
measurements of the second-order moment of the coarse-grained density
$\rho_r$ obtained by integrating the phase-space density
$f(\bm x,\bm v,t)$ with respect to velocities and over space in boxes
of length $r$. This is the two-dimensional version of
Fig.~\ref{fig:1D_M2_convergence} and the statistics of
$\langle\rho_r^2\rangle$ have a very similar behavior as in the
one-dimensional case.  Here $\St=0.5$,
$V_{\rm max} =3.9\,u_{\rm rms}$, and $N_v$ is varied from 9 to 21. One
clearly observes that the statistics obtained from the lattice method
converges to that obtained from Lagrangian simulations.
		
The interplay between the choices of $N_v$ and of $V_{\rm max}$
requires some further comments. On the one hand the method converges
when both $\Delta v = 2\,V_{\rm max}/N_v \to 0$ and
$V_{\rm max}\to\infty$. On the other hand, the computational cost is
$\propto N_v^2$.  One can thus wonder if for a fixed cost there is an
optimal choice of $V_{\rm max}$ that minimizes the error obtained with
the lattice method. Focusing again on the second-order statistics of
the particle mass distribution, we have measured the average with
respect to $r$ of the error made on the density moment
$\langle\rho_r^2\rangle$ defined as
\begin{equation}
 \mathcal{E}(N_v,V_{\rm max})  = \overline{ \langle(\rho_r^{\rm E})^2\rangle
  - \langle(\rho_r^{\rm L})^2\rangle } / \overline{\langle(\rho_r^{\rm
    L})^2\rangle},
  \label{eq:defGlobErrM2}
\end{equation}
where
\begin{equation}
  \overline{ f(r) } = \frac{1}{L^2} \int_0^L |f(r)|\,r\,\mathrm{d}r.
\end{equation}

\begin{figure}[ht]
  \centerline{\includegraphics[width=0.9\columnwidth]{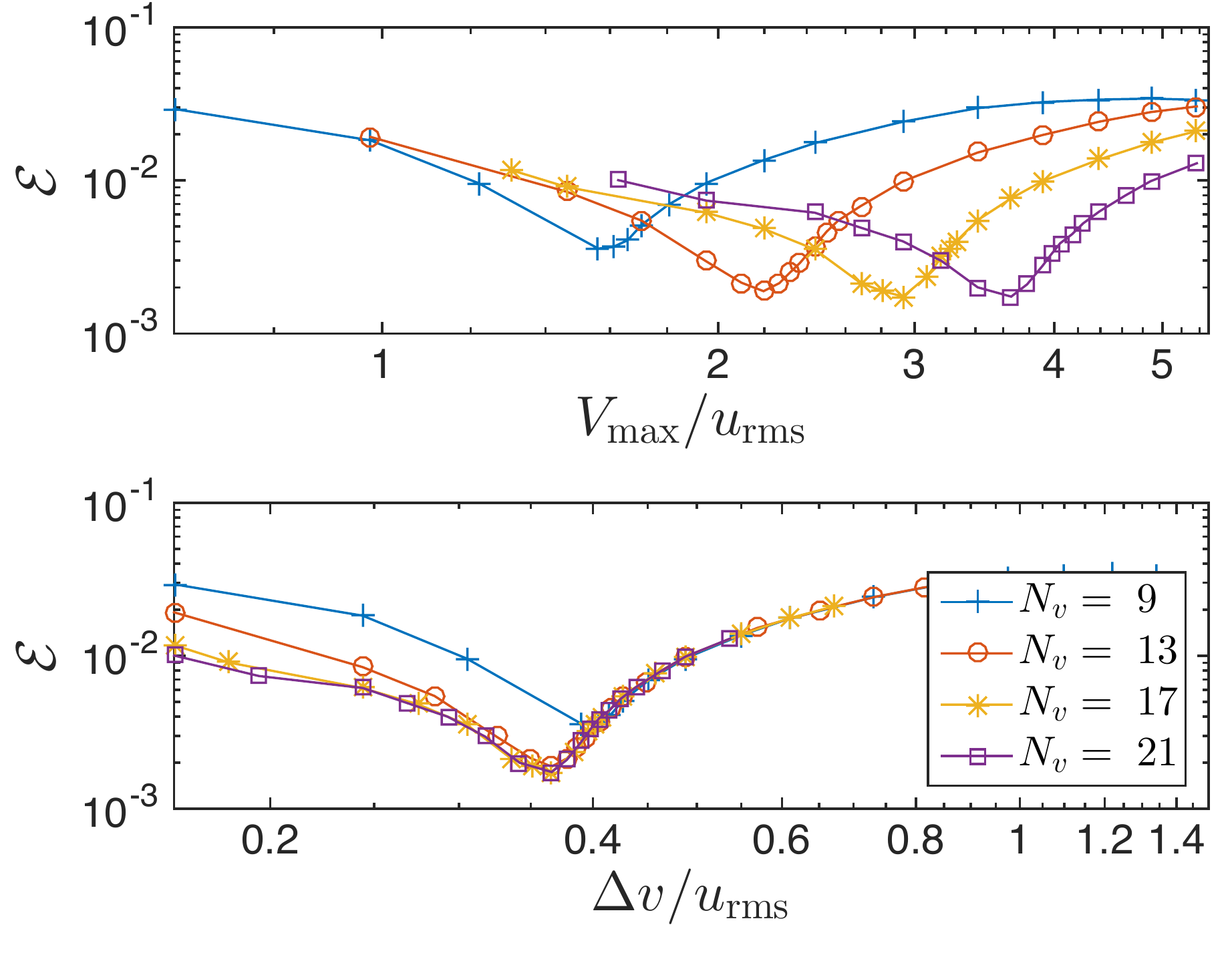}}
  \caption{Distance-averaged error $\mathcal{E}$ of the second-moment
    of the mass density as defined in (\ref{eq:defGlobErrM2}) as a
    function of the maximal velocity $V_{\rm max}$ (top) and of the
    velocity grid spacing $\Delta v = 2\,V_{\rm max}/ (N_v-1)$ (bottom)
    for various values of the velocity resolution $N_v^2$ and for
    $\St \approx 0.5$.}
  \label{fig:2D_structF}
\end{figure}
Figure~\ref{fig:2D_structF} (top) represents this quantity as a
function of $V_{\rm max}$ for different values of the cost $N_v^2$.  One
clearly observes that there is indeed for a fixed $N_v$ a specific
choice of $V_{\rm max}$ where the error is minimal. The optimal value
of the maximal velocity increases with $N_v$. On the right of the
minimum, the error is in principle dominated by a $\Delta v$ too
large. This is confirmed by the collapse of the various curves on the
right of their minima that can be seen in the bottom of
Fig.~\ref{fig:2D_structF} where $\mathcal{E}$ is represented as a
function of $\Delta v$. In the left of the minimum, the error should
be dominated by a too small value of $V_{\rm max}$. One can indeed
guess an asymptotic collapse for $V_{\rm max}\ll u_{\rm rms}$ on the
upper panel of Fig.~\ref{fig:2D_structF}, or equivalently, the fact
that the curves separate from each other at small values of $\Delta v$
in the lower panel.

The value of the error at the optimal $V_{\rm max}$ decreases from
$N_v=9$ to $N_v=13$ but then seems to saturate (or to decrease only
very slowly) at higher values of $N_v$. One cannot exclude that this
behavior corresponds to a logarithmic convergence of $\mathcal{E}$
when $N_v\to\infty$. This slow dependence is also visible in the
bottom panel where the collapse of the various curves seems to extend
weakly on the left-hand side of the minima for $N_v = 13$, $17$ and
$21$. Accordingly, a small difference in $N_v$ is not enough to
decrease significantly the error. In the specific case considered (for
$\St\approx 0.5$ in the direct cascade), the resulting optimal choice
seems to be $N_v = 13$ with $V_{\rm max} = 2.25\,u_{\rm rms}$, which
leads to a relative error $ ~ 10^{-3}$.

\subsubsection{Inverse energy cascade}

To complete this study we have also tested the proposed latice method
in a two-dimensional turbulent flow in the inverse kinetic energy
cascade regime. The stochastic forcing is now acting at small scales
($400\le|\bm k|\le 405$) and we made use of hyper-viscosity (here
eighth power of the Laplacian) in order to truncate the direct
enstrophy cascade. The kinetic energy accumulated at large scales is
again removed using a linear friction in the Navier--Stokes equation
(\ref{eq:ns2D}). The particle Stokes number is now defined as
$\St = \tau_{\rm p} / \tau_L$ using the large-eddy turnover time
$\tau_L = L / u_{\rm rms}$ since small-scale statistics are dominated
by forcing and are thus irrelevant. The flow is integrated with a
resolution of $2048^2$ gridpoints while the lattice-particle method is
applied for $\St\approx 0.1$ on a coarser grid with $512^2$ points.

\begin{figure}[ht]
  \centerline{\includegraphics[width=0.85\columnwidth, height=0.85\columnwidth]{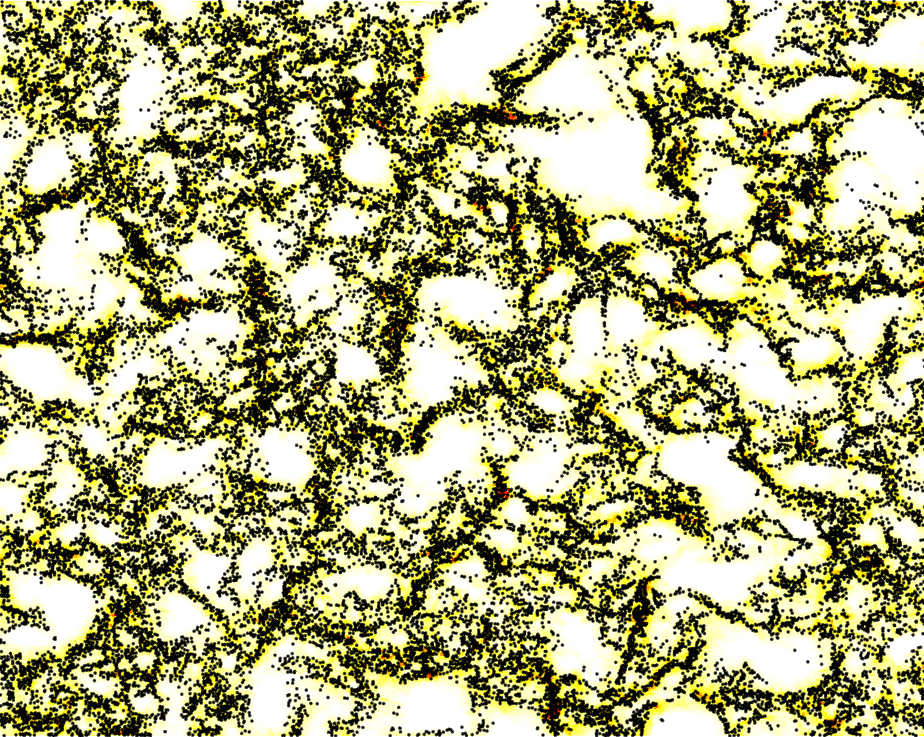}}
  \caption{Snapshot of the position of Lagrangian particles (black
    dots), together with the density field obtained from the lattice
    method (colored background, from white: low densities to red: high
    densities) for $\St\approx 0.1$ in the inverse energy cascade. The
    fluid flow was integrated using a resolution of $2048^2$ and
    lattice simulations were performed with $512^2$ spatial
    gridpoints associated to $9^2$ discrete velocities.}
  \label{fig:inv_snap}
\end{figure}
\begin{figure}[ht]
  \centerline{\includegraphics[width=0.85\columnwidth]{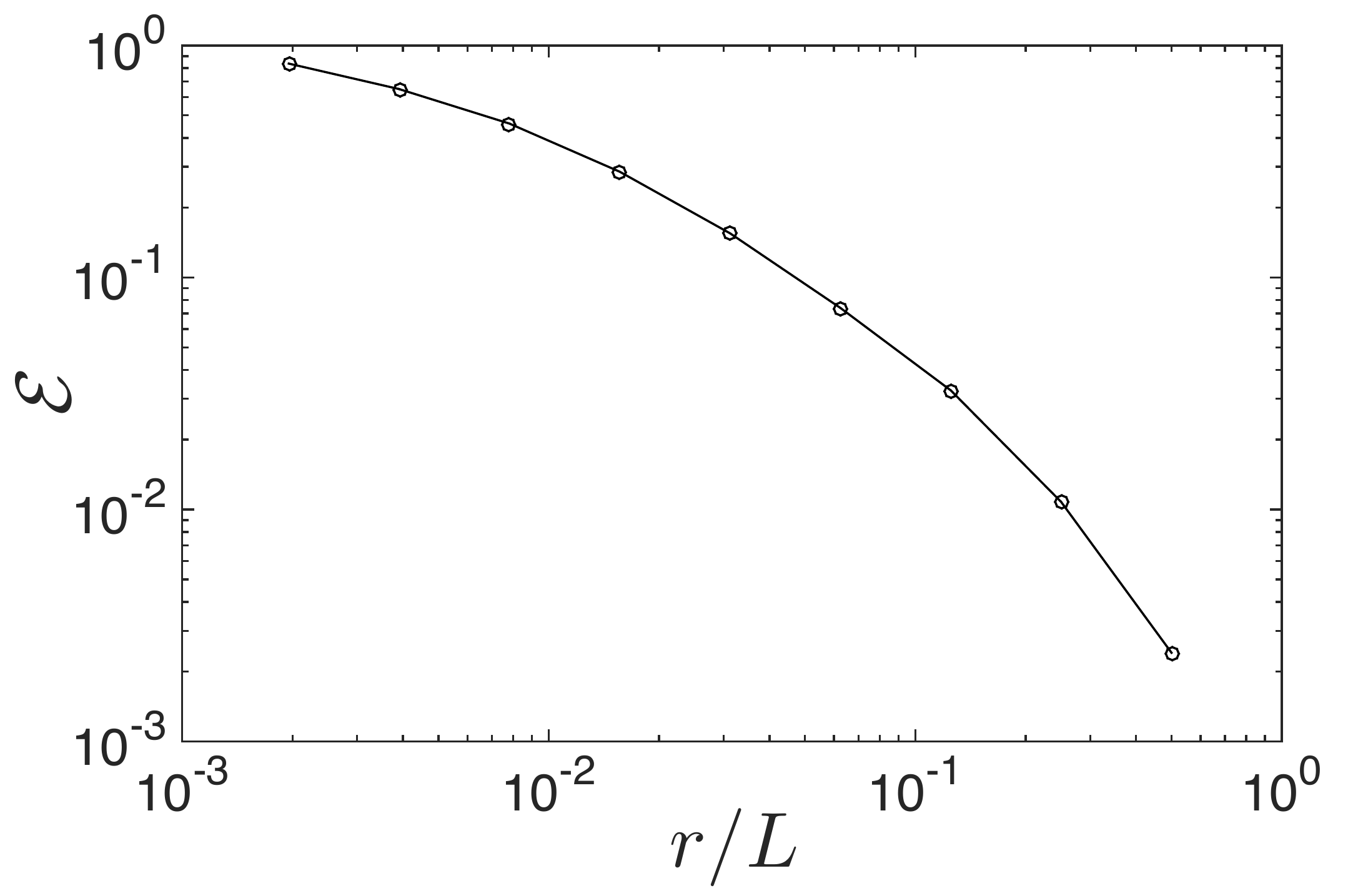}}
  \caption{Relative error $\mathcal{E}$ of the second-order moment of
    the coarse-grained density $\rho_r$ as a function of $r$ for
    $\St=0.5$. The lattice-particle method was here used with $512^2$
    position gridpoints and $9^2$ velocity gridpoints.}
  \label{fig:2D_INV_M2}
\end{figure}

Figure~\ref{fig:inv_snap} shows that the lattice-particle method is
able to reproduce the main qualitative features of the particle
spatial distribution at scales within the inertial range of the
inverse energy cascade. This is confirmed in Fig.~\ref{fig:2D_INV_M2}
which represents the relative error $\mathcal{E}$ defined in
Eq.~\eqref{eq:defGlobErrM2} of the second-order moment
$\langle \rho_r^2 \rangle$ of the density $\rho_r$ coarse-grained over
a scale $r$. The Lagrangian integration was performed with
$2\times 10^7$ particles (with no physical diffusion) and the lattice
method on a $512^2$ spatial grid with $N_v^2 = 9^2$ discrete values of
the particle velocity. One clearly observes that the error decreases
at the largest scales of the flow.

%% SECTION %%

\section{Conclusions}

We have presented a new Eulerian numerical method to model the
dynamics of inertial particles suspended in unsteady flow. This
lattice-particle method is based on the discretization in the
position-velocity phase space of the evolution equation for the
particle distribution.  The spatial grid is chosen such that particles
with a given discrete velocity hop by an integer number of gridpoints
during one time step, an idea close to that used in lattice-Boltzmann
schemes. We have shown that the model reproduces the correct dynamical
and statistical properties of the particles, even with a reasonably
small amount of velocity gridpoints. Some deviations from Lagrangian
measurements are nevertheless observed at small scales. We obtained
evidence that they are due to numerical diffusivity acting in the
space of velocities. The proposed method is anyway intended to
describe large scales where such deviations disappear. It might hence
be a suitable candidate for developing large-eddy models for particle
dynamics. Indeed, as equation~\eqref{eq:conslaw} is linear in $f$,
some techniques of subgrid modeling used in scalar turbulent transport
\citep[see, for example,][]{girimaji1996analysis} could be adapted.

Our approach consists in always imposing the same mesh for particle
velocities, independently of the spatial position and of the local
value of the fluid velocity. This is well-adapted for particles with a
large Stokes number.  Their velocity experiences small fluctuations
and is generally poorly connected to that of the fluid. In addition,
the method is accurate at the largest scales and can hence catch the
structures appearing in the spatial and velocity distributions of
large-Stokes-number particles.  Such considerations indicate that the
proposed lattice-particle method is suitable for modeling particles
with a sufficiently strong inertia.  Conversely, particles with a weak
inertia develop fine-scale structures in their distribution.  They
result from tiny departures of their velocity from that of the
fluid. Our method, applied with a fixed velocity resolution, might not
be able to catch such deviations. However, a more suitable idea for
this case is to use a variation of our approach where, instead of a
full resolution of the particle velocity, one considers its difference
with that of the fluid.  This would of course require changing scheme
for integrating advection.

\section*{Acknowledgments}
This study benefited from fruitful discussions with M.~Gorokhovski,
H.~Homann, S.~Musacchio.
The research leading to these results has received funding from the
Agence Nationale de la Recherche (Programme Blanc ANR-12-BS09-011-04)
and from the European Research Council under the European Community's
Seventh Framework Program (FP7/2007-2013 Grant Agreement
no.~240579). Numerical simulations were performed on the
``m\'{e}socentre de calcul SIGAMM''.

								%% BIBLIOGRAPHY %%

%\bibliographystyle{model5-names}
\bibliographystyle{elsarticle-harv}
\bibliography{./biblio}

\end{document}